\documentclass[journal]{IEEEtran}
\usepackage{amsmath,amsfonts}
\usepackage{algorithm}
\usepackage{algpseudocode}
\usepackage{array}
\usepackage[caption=false,font=normalsize,labelfont=sf,textfont=sf]{subfig}
\usepackage{textcomp}
\usepackage{stfloats}
\usepackage{url}
\usepackage{verbatim}
\usepackage{graphicx}
\usepackage{cite}
\usepackage{xcolor}
\usepackage{mathtools}
\usepackage{amssymb}
\usepackage{amsthm}

\algrenewcommand\algorithmicrequire{\textbf{Input:}}
\algrenewcommand\algorithmicensure{\textbf{Output:}}

\newcommand{\transpose}{\mathsf{T}}
\newcommand{\norm}[1]{\left\lVert #1 \right\rVert}
\renewcommand{\thesection}{\Roman{section}}

\usepackage{booktabs}
\usepackage{multirow}
\usepackage{color}
\usepackage{multicol}
\usepackage[hidelinks]{hyperref}

\begin{document}
\title{
Weighted Covariance Intersection for Range-based Distributed Cooperative Localization of Multi-Vehicle Systems}

\author{Chenxin~Tu,
        Xiaowei~Cui,
        Gang~Liu,
        and Mingquan Lu 
\thanks{Copyright (c) 20xx IEEE. Personal use of this material is permitted. However, permission to use this material for any other purposes must be obtained from the IEEE by sending a request to pubs-permissions@ieee.org.}
\thanks{This work was supported in part by the National Key R\&D Program of China under Grant 2021YFA0716603 and in part by the National Natural Science Foundation of China under Grant U2233217. (\textit{Corresponding authors: Xiaowei Cui; Mingquan Lu}).}
\thanks{Chenxin Tu and Gang Liu are with the Department of Electronic Engineering, Tsinghua
University, Beijing 100084, China (e-mail: tcx22@mails.tsinghua.edu.cn; liu\_gang@tsinghua.edu.cn).}
\thanks{Xiaowei Cui and Mingquan Lu are with the Department of Electronic Engineering, State Key Laboratory of Space Network and Communication, Tsinghua University, Beijing 100084, China (e-mail:
cxw2005@tsinghua.edu.cn; lumq@tsinghua.edu.cn).}
}



\maketitle

\begin{abstract}
Cooperative localization enables autonomous navigation for multi-vehicle systems (MVS) in GNSS-denied environments. Among the available architectures, distributed cooperative localization (DCL) is attractive for its robustness and scalability in large-scale MVS. To address the challenge of untrackable state correlations between vehicles in a distributed framework, covariance intersection (CI) has been introduced as a means to fuse inter-vehicle measurements. However, existing studies typically treat CI as a plug-in technique, directly applying traditional optimization criteria and focusing mainly on simple two-dimensional (2D) scenarios. When extended to three-dimensional (3D) cooperative localization with higher-dimensional states and pronounced disparities in scale and observability among state components, traditional CI criteria fail to maintain balanced estimation performance across the full state, and some state components suffer substantial accuracy degradation. This limitation calls for task-oriented improvements to the CI fusion process. In this paper, we introduce a weighting mechanism, called weighted covariance intersection (WCI), to regulate the CI fusion process in 3D DCL. We further develop a concurrent fusion strategy for multiple distance measurements and design a dedicated weighting matrix based on inertial navigation system (INS) error propagation.
The method supports diverse MVS platforms, including unmanned aerial vehicle (UAV) swarms and autonomous ground vehicle (AGV) fleets, in complex 3D environments.
Simulation results show that the proposed WCI significantly improves cooperative localization performance over traditional CI, while the distributed framework offers clear advantages over the centralized counterpart in robustness and scalability.
\end{abstract}


\section{Introduction}
Autonomous vehicles, such as unmanned aerial vehicles (UAVs) and unmanned ground vehicles (UGVs), have found wide application in environmental monitoring, disaster rescue, and military operations \cite{asadzadeh2022uav, erdos2013experimental, ma2013simulation}. In recent years, multi-vehicle systems (MVS) have attracted increasing attention because their flexibility, cooperative capability, and robustness enable them to accomplish complex tasks efficiently \cite{scherer2015autonomous, alotaibi2019lsar}. Accurate positioning is a fundamental requirement for such systems. At present, the Global Navigation Satellite System (GNSS) remains one of the most widely used positioning solutions for MVS \cite{hacohen2020improved}. However, although GNSS can provide all-day, all-weather positioning outdoors, its signals can be severely blocked or become unavailable in indoor and underwater environments \cite{kaplan2017understanding}. Local positioning systems (LPSs), on the other hand, require predeployed infrastructure \cite{tu2025parameterized,tu2025decoupled}. In this context, cooperative localization has emerged as an attractive alternative, since collaboration among vehicles can improve both positioning accuracy and availability in GNSS-denied environments \cite{roumeliotis2002distributed,liu2024cooperative,van2015least}.
With the growing demand for MVS in transportation, autonomous driving, and electric vehicle fleet management, cooperative localization has become increasingly relevant to modern vehicular systems \cite{zhu2024energy, wang2025optimization, yuan2025vehicle}.

Since cooperative localization was first introduced in \cite{kurazume1994cooperative}, substantial progress has been made over the past three decades. In a typical cooperative-localization system, vehicles obtain relative measurements, exchange information through vehicle-to-vehicle (V2V) communication, and fuse local and cooperative information so that well-localized vehicles can assist others in improving their global localization performance \cite{li2024vehicles,zhu2020incentive}. Among the available approaches, filter-based cooperative localization is particularly attractive because it recursively estimates vehicle states, including position and attitude, and thus directly supports autonomous navigation.
Since each vehicle uses information received from other vehicles to estimate its own state, inter-vehicle correlations naturally arise during cooperation and must be accounted for in the update process. These correlations originate from inter-vehicle measurement updates and subsequent propagation through prediction. We refer to the state update based on inter-vehicle measurements as the correlated update. Depending on whether the computation is carried out at a computing center or locally at each vehicle, cooperative localization can be categorized into centralized cooperative localization (CCL) and distributed cooperative localization (DCL) \cite{li2013cooperative}. The two architectures differ mainly in how they handle the evolving inter-vehicle correlations. CCL explicitly tracks these correlations by collecting all information at a computing center, but this comes at the price of high communication overhead and limited robustness and scalability. By contrast, DCL typically neither requires nor can afford explicit correlation tracking, which makes it more suitable for large-scale swarms and complex operating conditions. Its central challenge, however, is how to perform effective correlated updates when the inter-vehicle correlations are unknown.

Covariance intersection (CI) was originally proposed as a distributed fusion method that can produce a consistent and conservative estimate under unknown correlations between data sources \cite{julier1997non}. It has since been widely used in recursive filtering problems involving unknown correlations between measurements and estimates such as cooperative perception \cite{lima2021data}. After Arambel introduced CI into cooperative localization \cite{arambel2001covariance}, CI and its variants were successfully adopted in a variety of cooperative-vehicle-localization studies. Carrillo and Li fused relative-pose-based estimates using CI and split covariance intersection (SCI), respectively \cite{li2013cooperative,li2013split,carrillo2013decentralized}. Pierre extended CI-based correlated updates to partial-observation scenarios and achieved DCL using only distance measurements \cite{pierre2018range}. Chang developed a CI-based cooperative localization scheme that decouples measurement and communication, thereby improving the robustness and scalability of DCL \cite{chang2021resilient}. Fang addressed cooperative localization with heterogeneous error sources by using an adaptive cubature split covariance intersection filter \cite{fang2021adaptive}. 

\begin{figure}
    \centering
    \includegraphics[width=\linewidth]{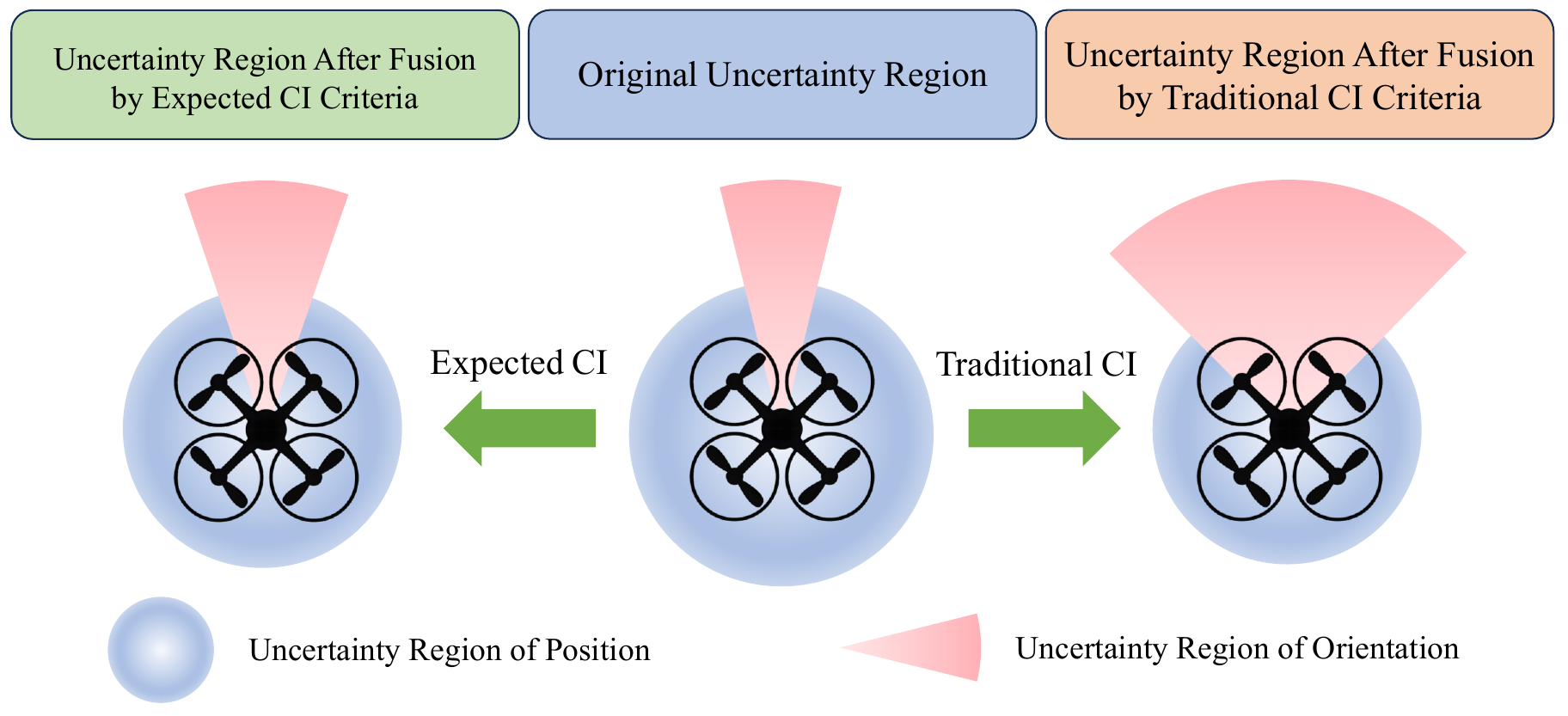}
    \caption{Illustrative example used to motivate the need for balanced fusion in 3D cooperative localization. For simplicity, only a 2D position-attitude subspace of the full state is shown. Under traditional CI criteria, fusion may reduce position error at the expense of substantially increasing attitude error. The goal is therefore a more balanced fusion that improves position accuracy without severely degrading attitude estimation.
    }
    \label{fig1}
\end{figure}

Despite this substantial body of work, two limitations remain. First, most existing studies treat CI as a plug-in fusion tool and directly adopt traditional trace-based \cite{carrillo2013decentralized,chang2021resilient} or determinant-based \cite{li2013cooperative,li2013split,pierre2018range,fang2021adaptive} criteria, without examining how the choice of criterion affects the cooperative-localization task itself. Second, these studies mainly focus on MVS operating in two-dimensional (2D) scenarios, where vehicles are usually modeled with a unicycle-like state of only three dimensions. By contrast, DCL for MVS operating in three-dimensional (3D) environments, such as UAV swarms, remains much less explored. 
In 3D scenarios, the vehicle state is higher dimensional, contains additional state components, and exhibits pronounced disparities in both scale and observability. Under such conditions, traditional CI criteria may fail to deliver balanced full-state fusion, and some state components may suffer substantial performance degradation. Fig.~\ref{fig1} provides an illustrative example of this issue.
This issue is critical because small-scale and weakly observable state components, such as attitude, play important roles alongside position in vehicle operation and the long-term stability of localization. When their estimation errors become large, autonomous navigation performance deteriorates markedly, and the position error can become highly susceptible to divergence, especially when observations are insufficient. Therefore, 3D cooperative localization requires a more suitable CI optimization criterion and a correspondingly improved fusion mechanism.

Weighting mechanisms are widely used to regulate data fusion because they provide flexibility in emphasizing quantities that are more relevant to a specific task \cite{kay1993fundamentals}. Incorporating such a mechanism into CI is therefore a promising way to improve correlated data fusion in the presence of large disparities in scale and observability among state components. More importantly, weighting enables the CI criterion to become task-oriented. In many practical problems, different state components contribute differently to the final objective, and a generic mean squared error (MSE) criterion cannot fully reflect this difference. In contrast, weighted mean squared error (WMSE) provides a more meaningful objective in such settings \cite{eldar2008universal}. Although the concept of weighted covariance intersection (WCI) was already mentioned in \cite{arambel2001covariance}, it was used mainly heuristically to improve selected state components, without clarifying why weighting is fundamentally needed in scenarios with large inter-state disparities or what optimization objective it should represent. These limitations have hindered the broader use of WCI in recursive filtering problems such as cooperative localization. In this paper, we address both issues and integrate WCI into DCL for 3D scenarios in a principled way.

To achieve balanced full-state estimation in range-based 3D distributed cooperative localization, we introduce WCI into the correlated-update process. Specifically, we develop a concurrent fusion strategy for multiple distance measurements that is compatible with the characteristics of CI fusion, and we design a dedicated weighting matrix based on inertial navigation system (INS) error propagation. Extensive simulations are conducted to validate the effectiveness of the proposed method. The results show that WCI can substantially improve the balance and overall quality of state estimation in 3D cooperative localization compared with traditional CI, while the distributed framework retains clear advantages over the centralized framework in robustness and scalability.

The remainder of this paper is organized as follows. In Section~\textrm{II}, we introduce the setup of the MVS under study, including the system and state composition, as well as range measurement models. In Section~\textrm{III}, we introduce the overall distributed cooperative localization framework and focus on the correlated update process. Section~\textrm{IV} consists of three parts: a review of CI and WCI, including an analysis of the limitations of traditional CI criteria under scale and observability disparities; a concurrent fusion strategy for multiple distance measurements; and a weighting matrix design based on the error propagation rule of INS. In Section~\textrm{V}, we present extensive simulations to validate the effectiveness of the proposed DCL method and demonstrate its superiority over the CCL method. Finally, the conclusions are summarized in Section~\textrm{VI}.

The main notation used in the paper is summarized in Table~\ref{tab1}.
\begin{table}[!t]
\centering
\caption{Summary of main notation.}
\setlength{\tabcolsep}{5pt}
\begin{tabular}{p{2.1cm} p{6cm}}
\hline
\textbf{Notation} & \textbf{Description} \\
\hline
lowercase $x$ & scalar \\
bold lowercase $\boldsymbol{x}$ & vector \\
bold uppercase $\boldsymbol{X}$ & matrix \\
$\left\lVert \boldsymbol{x} \right\rVert$ & Euclidean norm of a vector \\
$\mathbb{E}[\cdot]$ & expectation operator \\
$\operatorname{tr}(\cdot)$ & trace operator \\
$\operatorname{det}(\cdot)$ & determinant operator \\
$\operatorname{diag}(\cdot)$ & diagonal matrix with the entries inside \\
$(\cdot)^{\transpose}$ & transpose operator \\
$\hat{(\cdot)}$ & noise-contaminated estimate \\
$\tilde{(\cdot)}$ & noise-contaminated measurement \\
$(\cdot)\times$ & antisymmetric matrix constructed from a vector \\
$\times$ & cross-product operator \\
$\succeq$ & greater than or equal to in the L\"{o}wner partial order; e.g., $\boldsymbol{A} \succeq \boldsymbol{B}$ indicates that $\boldsymbol{A}-\boldsymbol{B}$ is positive semidefinite \\
$\boldsymbol{I}_M$ & $M \times M$ identity matrix \\
$\boldsymbol{0}_{M\times N}$ & $M \times N$ zero vector or matrix \\
$\mathbb{R}^{M\times1}$ & $M \times 1$  real-valued vector \\
\hline
\end{tabular}
\label{tab1}
\end{table}

\section{System Setting}

\begin{figure}
    \centering
    \includegraphics[width=\linewidth]{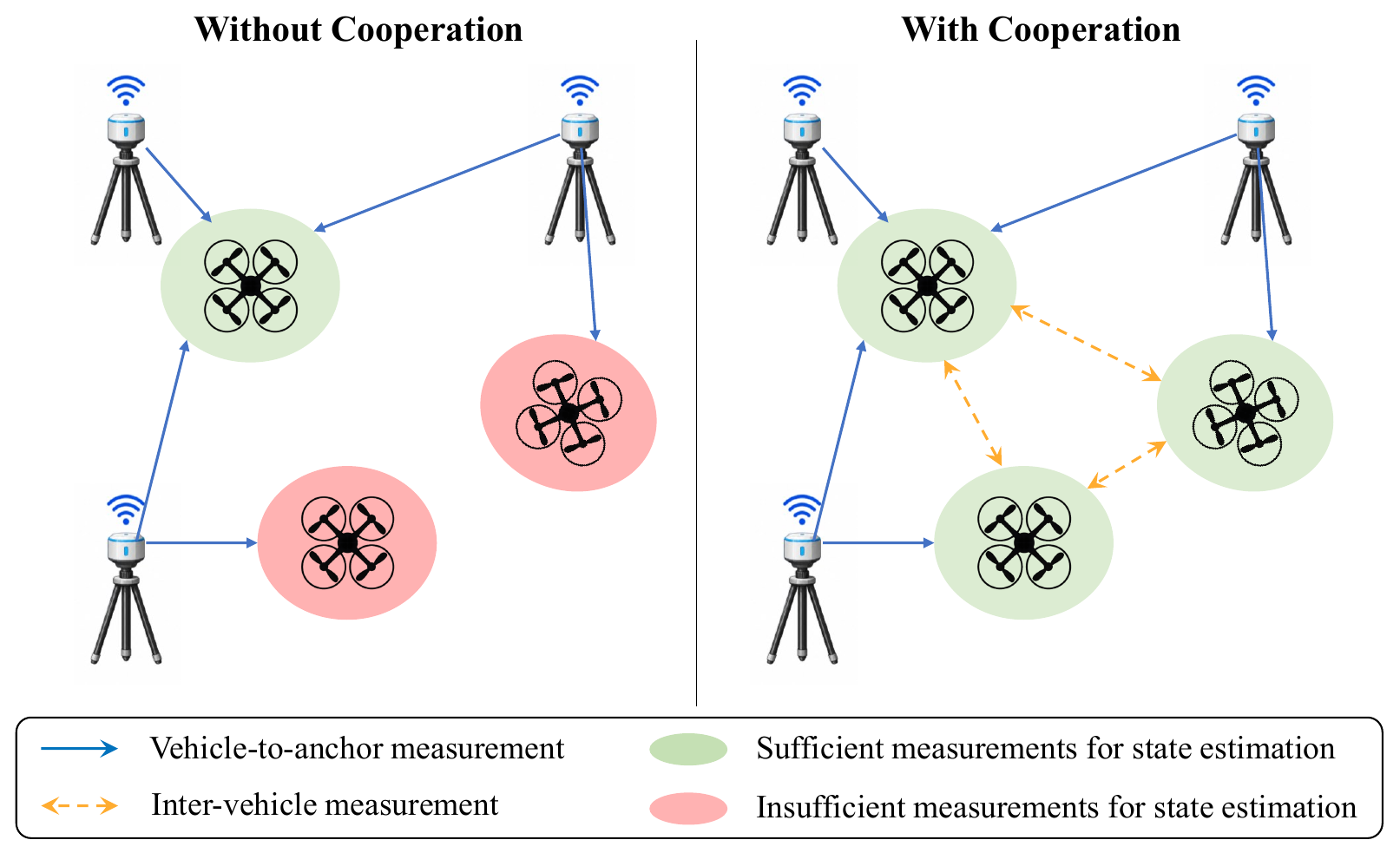}
    \caption{Schematic of a range-based cooperative localization system, where vehicles with insufficient anchor measurements can achieve accurate state estimation through inter-vehicle cooperation.}
    \label{fig2}
\end{figure}

We consider a range-based cooperative localization system composed of an MVS with $N_v$ vehicles moving in 3D space and $N_a$ fixed anchors with known positions, as illustrated in Fig.~\ref{fig2}. Owing to communication-range limitations and environmental occlusions, some vehicles may not obtain sufficient anchor measurements for reliable state estimation. In such cases, cooperative localization allows them to improve their state estimates through inter-vehicle ranging and information exchange. Each vehicle is equipped with an onboard inertial measurement unit (IMU) for proprioceptive motion sensing. Both the vehicles and anchors are also equipped with rigidly mounted ranging modules, allowing each vehicle to measure distances to neighboring vehicles and anchors within its communication range\footnote{In this study, anchors are assumed to function solely as signal transmitters, with measurements carried out only on the vehicle side.}. In addition, both the vehicles and anchors are equipped with communication modules for information exchange.

\begin{figure}
    \centering
    \includegraphics[width=\linewidth]{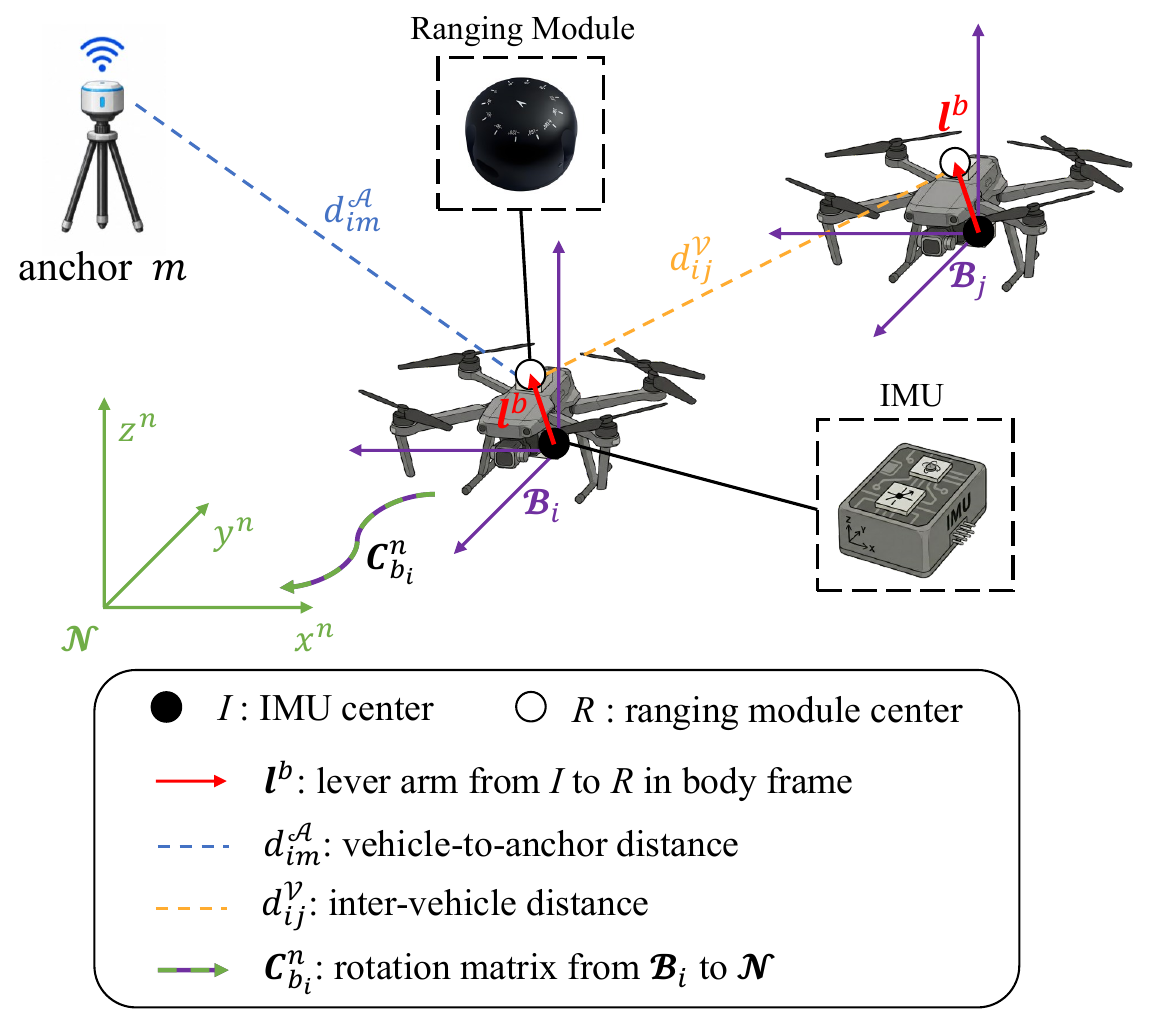}
    \caption{
    Geometric relationships among the main frames and vectors in the system model. $\mathcal{N}$ denotes the navigation frame, and $\mathcal{B}_i$ and $\mathcal{B}_j$ denote the body-fixed frames of vehicles $i$ and $j$, respectively. The lever arm $\boldsymbol{l}^b$ connects the IMU center (superscript $I$) to the ranging module (superscript $R$) on each vehicle. $d_{ij}^{\mathcal{V}}$ denotes the inter-vehicle distance between vehicles $i$ and $j$, and $d_{im}^{\mathcal{A}}$ denotes the vehicle-to-anchor distance between vehicle $i$ and anchor $m$.
    }
    \label{fig3}
\end{figure}

Since the IMU is used as the proprioceptive sensor, the state of vehicle $i$ ($i=1,\ldots,N_v$) is defined as follows:
\begin{equation}
    \boldsymbol{x}_i = \left[ 
    \left( \prescript{I}{}{\boldsymbol{p}}_i^{n} \right)^{\mathsf{T}}, 
    \left( \prescript{I}{}{\boldsymbol{v}}_i^{n} \right)^{\mathsf{T}}, 
    \left( \boldsymbol{q}_i^{n} \right)^{\mathsf{T}},
    \boldsymbol{b \scriptstyle g}_i^{\mathsf{T}},
    \boldsymbol{b \scriptstyle a}_i^{\mathsf{T}}
    \right]^\mathsf{T},
\label{eq1}
\end{equation}
where $\prescript{I}{}{\boldsymbol{p}}_i^{n}$, $\prescript{I}{}{\boldsymbol{v}}_i^{n}$, and 
$\boldsymbol{q}_i^{n}$ denote the position, velocity, and attitude of vehicle $i$ in the navigation frame $\mathcal{N}$, with the attitude represented by a quaternion. $\boldsymbol{b \scriptstyle g}_i$ and $\boldsymbol{b \scriptstyle a}_i$ denote the gyroscope and accelerometer biases of the IMU, respectively. Here, the left superscript $I$ indicates that the position is referenced to the IMU center. In practice, the IMU and the ranging module are generally not co-located, so a lever arm exists between them. We assume that all vehicles share an identical lever arm, denoted by $\boldsymbol{l}^{b}$ in the body-fixed frame $\mathcal{B}$. 
The main geometric relationships are shown in Fig.~\ref{fig3}.

The position of the ranging module on vehicle $i$ can then be expressed as
\begin{equation}
    \prescript{R}{}{\boldsymbol{p}_i^n}= \prescript{I}{}{\boldsymbol{p}_{i}^n} + \boldsymbol{C}_{b_i}^{n} \boldsymbol{l}^{b},
\label{eq2}
\end{equation}
where $\boldsymbol{C}_{b_i}^{n}$ denotes the rotation matrix from frame $\mathcal{B}_i$ to frame $\mathcal{N}$, and the left superscript $R$ indicates that the position corresponds to the ranging module on the vehicle. The relative distances between ranging modules can be written as 
\begin{equation}
     d_{ij}^{\mathcal{V}} = \lVert \prescript{R}{}{\boldsymbol{p}_{i}^{n}} - \prescript{R}{}{\boldsymbol{p}_{j}^{n}}   \rVert \qquad
    d_{im}^{\mathcal{A}} = \lVert \prescript{R}{}{\boldsymbol{p}_{i}^{n}} - \prescript{A}{}{\boldsymbol{p}_m^{n}} \rVert,
\label{eq3}
\end{equation}
where $d_{ij}^{\mathcal{V}}$ is the distance between the ranging modules of vehicles $i$ and $j$, while $d_{im}^{\mathcal{A}}$ is the distance between the ranging module of vehicle $i$ and anchor $m$, whose position in frame $\mathcal{N}$ is denoted by $\prescript{A}{}{\boldsymbol{p}_m^{n}}$. The superscripts $\mathcal{V}$ and $\mathcal{A}$ distinguish inter-vehicle distances from vehicle-to-anchor distances. The distance measurements are modeled as 
\begin{equation}
     {\tilde{d}_{ij}^{\mathcal{V}}} = {{d}_{ij}^{\mathcal{V}}} + n_{d_{ij}^{\mathcal{V}}} \qquad
    {\tilde{d}_{im}^{\mathcal{A}}} = {{d}_{im}^{\mathcal{A}}} + n_{d_{im}^{\mathcal{A}}},
\label{eq4}
\end{equation}
where $n_{d_{ij}^{\mathcal{V}}}$ and $n_{d_{im}^{\mathcal{A}}}$ are assumed to be independent and identically distributed zero-mean Gaussian random variables with variance $\sigma_r^2$.

In IMU-based state estimation, system error propagation and state update are typically performed using the error-state formulation rather than the nominal state directly \cite{sola2017quaternion}. This treatment mitigates the redundancy of quaternion-based attitude representation and alleviates the associated nonlinearities. We define the error state of vehicle $i$ as follows:
\begin{equation}
    \delta \boldsymbol{x}_i = \left[ 
    \left( \prescript{I}{}{\delta\boldsymbol{p}}_i^{n} \right)^{\mathsf{T}}, 
    \left( \prescript{I}{}{\delta\boldsymbol{v}}_i^{n} \right)^{\mathsf{T}}, 
   (\boldsymbol{\phi}^n_i)^{\mathsf{T}},
    \left( \delta \boldsymbol{b \scriptstyle g}_i \right)^{\mathsf{T}},
    \left( \delta \boldsymbol{b \scriptstyle a}_i \right)^{\mathsf{T}}
    \right]^\mathsf{T}.
\label{eq5}
\end{equation}
Here, the phi-angle model $\boldsymbol{\phi}^n_i$ is adopted to represent the attitude error \cite{niu2021wheel}.

\section{Distributed Cooperative Localization Framework}
\begin{figure}
    \centering
    \includegraphics[width=\linewidth]{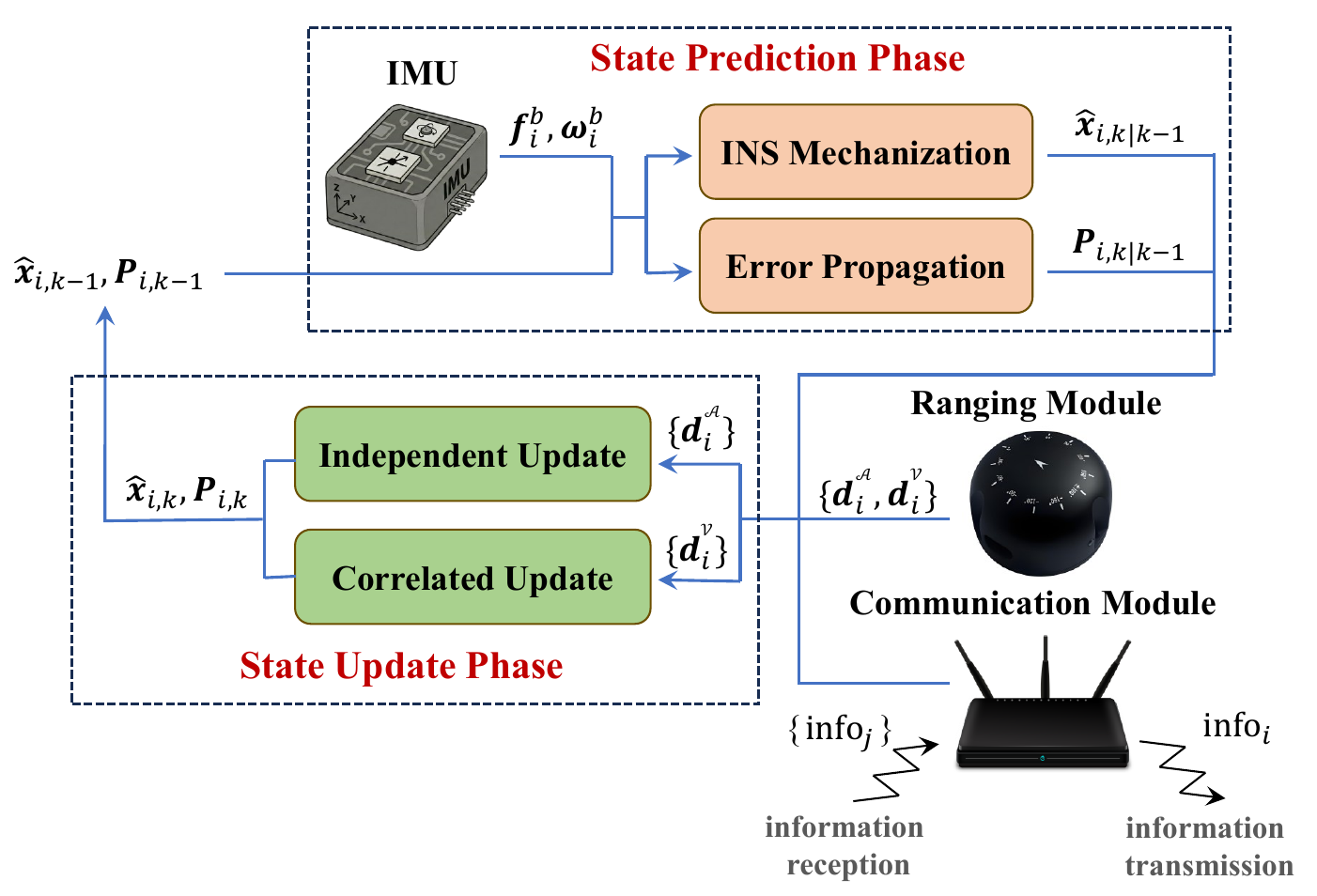}
    \caption{Range-based distributed cooperative localization framework for 3D scenarios. Here $\{ \boldsymbol{d}_i^{\mathcal{V}} \}$ and $\{ \boldsymbol{d}_i^{\mathcal{A}} \}$ denote the sets of distance measurements from vehicle $i$ to other vehicles and anchors, respectively. $\text{info}_i$ denotes the information broadcast by vehicle $i$, and $\{ \text{info}_j\}$ denotes the information received from other vehicles.}
\label{fig4}
\end{figure}

In this section, we introduce a range-based distributed cooperative localization framework for 3D scenarios. As illustrated in Fig.~\ref{fig4}, we use the state estimation process of vehicle $i$ at time step $k$ to explain the general DCL procedure.

In distributed cooperative localization, each vehicle estimates only its own state and maintains only its own state covariance matrix. The estimation process can be divided into two phases: state prediction and state update. During the prediction phase, the vehicle uses onboard IMU outputs to propagate the state and its covariance. During the update phase, it uses distance measurements to anchors and neighboring vehicles, together with information received from neighboring vehicles, to correct its state through data fusion. Accordingly, the distance measurements can be divided into two categories: vehicle-to-anchor measurements and inter-vehicle measurements. These two types require different update strategies.

\subsection{State Prediction Phase}
The state prediction phase includes the INS mechanization and the associated error propagation, which have been extensively studied and are not the primary focus of this paper. Therefore, we do not discuss these aspects in detail and refer readers to \cite{titterton2004strapdown, sola2017quaternion} for further information. It is worth emphasizing that, during prediction, state components other than position, such as velocity, attitude, and IMU biases, also affect the position estimation result. Therefore, in 3D cooperative localization, attention should not be limited to position accuracy alone; instead, all state components should be estimated in a balanced manner.
\subsection{State Update Phase}
\subsubsection{Observation Equation}
For the state update phase, we begin with the observation equations for distance measurements. Suppose the predicted state of vehicle $i$ is defined as follows
\begin{equation*}
    \hat{\boldsymbol{x}}_i = \left[ 
    \left( \prescript{I}{}{\hat{\boldsymbol{p}}}_i^{n} \right)^{\mathsf{T}}, 
    \left( \prescript{I}{}{\hat{\boldsymbol{v}}}_i^{n} \right)^{\mathsf{T}}, 
    \left( \hat{\boldsymbol{q}}_i^{n} \right)^{\mathsf{T}},
    \hat{\boldsymbol{b \scriptstyle g}}_i^{\mathsf{T}},
    \hat{\boldsymbol{b \scriptstyle a}}_i^{\mathsf{T}}
    \right]^\mathsf{T}.
\end{equation*}
The predicted position of the ranging module can be written as
\begin{equation}
\begin{split}
    \prescript{R}{}{\hat{\boldsymbol{p}}_{i}^{n}} &= \prescript{I}{}{\hat{\boldsymbol{p}}_{i}^{n}} + \hat{\boldsymbol{C}}_{b_i}^{n} \boldsymbol{l}^{b} \\
    &\approx (\prescript{I}{}{\boldsymbol{p}_{i}^{n}} + { 
 \prescript{I}{}{\delta \boldsymbol{p}_{i}^{n}}}) + \left[\boldsymbol{I}-({\boldsymbol{\phi}}_i^{n} \times) \right] \boldsymbol{C}_{b_i}^n \boldsymbol{l}^{b} \\
     &= \prescript{R}{}{\boldsymbol{p}_{i}^{n}} + { 
 \prescript{I}{}{\delta \boldsymbol{p}_{i}^{n}}} + [({\boldsymbol{C}}_{b_i}^n\boldsymbol{l}^{b})\times] \boldsymbol{\phi}_i^{n}.
\end{split}
\label{eq6}
\end{equation}
Then, the predicted distance between vehicles $i$ and $j$, as well as that between vehicle $i$ and anchor $m$, can be expressed as
\begin{equation}
\begin{aligned}
 {\hat{d}_{ij}^{\mathcal{V}}} = \ &
    \norm{\prescript{R}{}{\hat{\boldsymbol{p}}_{i}^{n}} - \prescript{R}{}{\hat{\boldsymbol{p}}_{j}^{n}}} \\
    \approx \ & \norm{\prescript{R}{}{\boldsymbol{p}_{i}^{n}} - \prescript{R}{}{\boldsymbol{p}_{j}^{n}}} +
    (\hat{\boldsymbol{r}}_{ij}^{\mathcal{V}})^{\transpose} \left( 
    \prescript{I}{}{\delta{\boldsymbol{p}}_{i}^{n}} -
    \prescript{I}{}{\delta{\boldsymbol{p}}_{j}^{n}}  \right) \\
    + \ &  (\hat{\boldsymbol{r}}_{ij}^{\mathcal{V}})^{\transpose} \left[ \left(  \hat{\boldsymbol{C}}_{b_i}^{n} \boldsymbol{l}^{b}\right)\times
     \right] \boldsymbol{\phi}_{i}^{n}
     - (\hat{\boldsymbol{r}}_{ij}^{\mathcal{V}})^{\transpose} \left[ \left(  \hat{\boldsymbol{C}}_{b_j}^{n} \boldsymbol{l}^{b}\right)\times
     \right] \boldsymbol{\phi}_{j}^{n} \\
    {\hat{d}_{im}^{\mathcal{A}}} = \ &
    \norm{\prescript{R}{}{\hat{\boldsymbol{p}}_{i}^{n}} - \prescript{A}{}{\boldsymbol{p}}_{m}^{n}} \\
    \approx \ & \norm{\prescript{R}{}{\boldsymbol{p}_{i}^{n}} - \prescript{A}{}{\boldsymbol{p}}_{m}^{n}} +
    (\hat{\boldsymbol{r}}_{im}^{\mathcal{A}})^{\transpose}  
    \prescript{I}{}{\delta{\boldsymbol{p}}_{i}^{n}} \\
    + \ &  (\hat{\boldsymbol{r}}_{im}^{\mathcal{A}})^{\transpose} \left[ \left(  \hat{\boldsymbol{C}}_{b_i}^{n} \boldsymbol{l}^{b}\right)\times
     \right] \boldsymbol{\phi}_{i}^{n},
\end{aligned}
\label{eq7}
\end{equation}
where $\hat{\boldsymbol{r}}_{ij}^{\mathcal{V}}$ and $\hat{\boldsymbol{r}}_{im}^{\mathcal{A}}$ denote the estimated unit direction vectors pointing from the ranging module of vehicle $j$ and anchor $m$ to the ranging module of vehicle $i$, respectively. Their explicit expressions are
\begin{equation}
    \hat{\boldsymbol{r}}_{ij}^{\mathcal{V}} = \frac{  \prescript{R}{}{\hat{\boldsymbol{p}}_{i}^{n}} - \prescript{R}{}{\hat{\boldsymbol{p}}_{j}^{n}}  }
    { \norm{\prescript{R}{}{\hat{\boldsymbol{p}}_{i}^{n}} - \prescript{R}{}{\hat{\boldsymbol{p}}_{j}^{n}}}} \qquad
    \hat{\boldsymbol{r}}_{im}^{\mathcal{A}} = \frac{ \prescript{R}{}{\hat{\boldsymbol{p}}_{i}^{n}} - \prescript{A}{}{{\boldsymbol{p}}_{m}^{n}}  }
    { \norm{\prescript{R}{}{\hat{\boldsymbol{p}}_{i}^{n}} - \prescript{A}{}{{\boldsymbol{p}}_{m}^{n}}}}.
\label{eq8}
\end{equation}
Since the error-state formulation is used for error propagation and state update, differential measurements are adopted instead of the original measurements. These differential measurements, namely the differences between the predicted distances and the actual distance measurements, can be expressed as
\begin{equation}
    \begin{aligned}
    \delta {d_{ij}^{\mathcal{V}}} &= {\hat{d}_{ij}^{\mathcal{V}}} - {\tilde{d}_{ij}^{\mathcal{V}}} = \boldsymbol{H}_{ij}^{\mathcal{V}} \delta \boldsymbol{x}_i + \boldsymbol{G}_{ij}^{\mathcal{V}} \boldsymbol{v}_{ij}^{\mathcal{V}} \\
     \delta {d_{im}^{\mathcal{A}}} &= {\hat{d}_{im}^{\mathcal{A}}} - {\tilde{d}_{im}^{\mathcal{A}}} = \boldsymbol{H}_{im}^{\mathcal{A}} \delta \boldsymbol{x}_i + v_{im}^{\mathcal{A}}
    \end{aligned},
    \label{eq9}
\end{equation}
Here, $\boldsymbol{v}_{ij}^{\mathcal{V}}$ and $v_{im}^{\mathcal{A}}$ denote the differential measurement noise terms, where $v_{im}^{\mathcal{A}} = -n_{d_{im}^{\mathcal{A}}}$. The expressions for $\boldsymbol{v}_{ij}^{\mathcal{V}}$ and its corresponding covariance matrix $\boldsymbol{R}_{ij}^{\mathcal{V}}$ are given by
\begin{equation}
    \boldsymbol{v}_{ij}^{\mathcal{V}} = 
    \left[ 
    \left( \prescript{I}{}{\delta{\boldsymbol{p}}_{j}^{n}} \right)^{\mathsf{T}}, 
    \left( \boldsymbol{\phi}_j^n \right)^{\transpose},
    n_{d_{ij}^{\mathcal{V}}}
    \right]^{\mathsf{T}},
    \label{eq10}
\end{equation}
\begin{equation}
\boldsymbol{R}_{ij}^{\mathcal{V}} =
\begin{bmatrix}
    \boldsymbol{P}_{{p}_j} & \boldsymbol{P}_{{p}_j  {\phi}_j} & \boldsymbol{0}_{3\times1}  \\
    \boldsymbol{P}_{{p}_j{\phi}_j}^{\transpose} & \boldsymbol{P}_{{\phi}_j} & \boldsymbol{0}_{3\times1}\\
    \boldsymbol{0}_{1\times3} & \boldsymbol{0}_{1\times3} & \sigma_r^2
\end{bmatrix}.
\label{eq11}
\end{equation}
Here, $\boldsymbol{P}_{{p}_j}$, $\boldsymbol{P}_{{\phi}_j}$, and $\boldsymbol{P}_{{p}_j  {\phi}_j}$ are the submatrices of the state covariance matrix of vehicle $j$, corresponding to the covariance matrices of its position and attitude, and their cross-covariance, respectively. In addition, the expressions for $\boldsymbol{H}_{ij}^{\mathcal{V}}$, $\boldsymbol{G}_{ij}^{\mathcal{V}}$, and $\boldsymbol{H}_{im}^{\mathcal{A}}$ are given by
\begin{equation}
    \begin{aligned}
    \boldsymbol{H}_{ij}^{\mathcal{V}} &=
    \left[  
    \left(\hat{\boldsymbol{r}}_{ij}^{\mathcal{V}} \right)^{\mathsf{T}},
    \boldsymbol{0}_{1\times3}, 
     \left(\hat{\boldsymbol{r}}_{ij}^{\mathcal{V}} \right)^{\mathsf{T}} \left[ \left(\hat{\boldsymbol{C}}_{b_i}^{n} \boldsymbol{l}^b \right) \times \right] ,
    \boldsymbol{0}_{1\times6}
    \right]\\
    \boldsymbol{G}_{ij}^{\mathcal{V}} &= 
    \left[  
    -\left(\hat{\boldsymbol{r}}_{ij}^{\mathcal{V}} \right)^{\mathsf{T}},
    -\left(\hat{\boldsymbol{r}}_{ij}^{\mathcal{V}} \right)^{\mathsf{T}}
    \left[ \left(\hat{\boldsymbol{C}}_{b_j}^{n} \boldsymbol{l}^b \right) \times\right],
    -1
    \right] \\
    \boldsymbol{H}_{im}^{\mathcal{A}} &= 
    \left[  \left( \hat{\boldsymbol{r}}_{im}^{\mathcal{A}} \right)^{\transpose},
    \boldsymbol{0}_{1\times3}, 
     \left( \hat{\boldsymbol{r}}_{im}^{\mathcal{A}} \right)^{\transpose} \left[ \left(\hat{\boldsymbol{C}}_{b_i}^{n} \boldsymbol{l}^b \right) \times \right] ,
    \boldsymbol{0}_{1\times6}
    \right]
    \end{aligned}.
\label{eq12}
\end{equation}

Assume that vehicle $i$ obtains distances to $M$ anchors and $N$ other vehicles for state update at time step $k$. Without loss of generality, let the indices of the anchors and neighboring vehicles be $1,\ldots,M$ and $1,\ldots,N$, respectively. Let the measurement vector formed by the differential distances to the $M$ anchors be denoted by $\boldsymbol{z}_{i}^{\mathcal{A}} = \left[ \delta d_{i1}^{\mathcal{A}}, \ldots, \delta d_{iM}^{\mathcal{A}} \right]^{\transpose}\in \mathbb{R}^{M\times1}$, with corresponding covariance matrix $\boldsymbol{R}_{i}^{\mathcal{A}} = \sigma_r^2 \boldsymbol{I}_{M}$. Let the differential distances to the $N$ other vehicles and their corresponding covariance matrices be denoted by $\left\{ \delta d_{ij}^{\mathcal{V}}, \boldsymbol{R}_{ij}^{\mathcal{V}}  \right\}$, $j=1,\ldots,N$. The predicted state and covariance matrix of vehicle $i$ before the update are given by $\left\{\hat{\boldsymbol{x}}_{i,k|k-1}, \boldsymbol{P}_{i,k|k-1}  \right\}$.
\subsubsection{Independent Update}
The differential distances between vehicle $i$ and anchors are independent of other vehicles' state uncertainties and can therefore be used directly for state update via EKF. Since the $M$ vehicle-to-anchor measurements are mutually independent, updating them sequentially as individual observations is equivalent to combining them into a single observation vector for concurrent update. We adopt the latter approach, and the corresponding update procedure is written as
\begin{gather}
    \boldsymbol{H}_{i}^{\mathcal{A}} = [{\boldsymbol{H}_{i1}^{\mathcal{A}}}^{\transpose}, \ldots, {\boldsymbol{H}_{iM}^{\mathcal{A}}}^{\transpose}]^{\transpose} \notag \\
    \boldsymbol{K} = \boldsymbol{P}_{i,k|k-1} {\boldsymbol{H}_{i}^{\mathcal{A}}}^{\transpose} \left( \boldsymbol{H}_{i}^{\mathcal{A}} \boldsymbol{P}_{i,k|k-1}  {\boldsymbol{H}_{i}^{\mathcal{A}}}^{\transpose} + \boldsymbol{R}_{i}^{\mathcal{A}}   \right)^{-1}  \notag  \\
    \delta \boldsymbol{x}_{i,k} = \boldsymbol{K} \boldsymbol{z}_{i}^{\mathcal{A}} \label{eq13} \\
    \boldsymbol{P}_{i,k} = 
    \left(\boldsymbol{I} - \boldsymbol{K}\boldsymbol{H}_{i}^{\mathcal{A}}\right) \boldsymbol{P}_{i,k|k-1} \notag.
\end{gather}
The estimated error state $\delta \boldsymbol{x}_{i,k}$ is then used to correct the nominal state estimate $\hat{\boldsymbol{x}}_{i,k|k-1}$, yielding the updated nominal state $\hat{\boldsymbol{x}}_{i,k}$.

\subsubsection{Correlated Update}

For inter-vehicle measurements, consider the distance measurement between vehicles $i$ and $j$ as an example, as shown in \eqref{eq9} and \eqref{eq10}. The differential distance $\delta d_{ij}^{\mathcal{V}}$ is used to update the state of vehicle $i$, while its noise term contains the position and attitude errors of vehicle $j$. Consequently, using this measurement introduces correlation between the state estimation errors of vehicles $i$ and $j$. In a distributed framework, tracking such cross-vehicle correlations is extremely challenging, or even infeasible. If these correlations are ignored and the estimates are incorrectly treated as independent, the resulting fusion may suffer from the ``data incest'' problem \cite{julier2017general}, which can lead to inconsistency or even divergence. Therefore, this work focuses on the correlated update problem, aiming to achieve effective state update with inter-vehicle distance measurements under unknown correlations while maintaining accurate and consistent 3D cooperative localization.

It should be emphasized that, although inter-vehicle range measurements directly couple vehicle positions and indirectly couple attitudes through the lever-arm effect, the resulting cross-vehicle correlations are not confined to these components. During the prediction step, the error-state transition propagates correlations across the entire state vector. In particular, the INS error propagation equations \cite{titterton2004strapdown} show that position error is coupled with velocity error, velocity error is coupled with attitude error and accelerometer bias, and attitude error is coupled with gyroscope bias. Consequently, once a correlated update introduces cross-vehicle correlation into any state component, such as position, subsequent prediction steps spread it to velocity, attitude, and all IMU bias states, eventually yielding full-state correlation between the two vehicles. This observation motivates the use of full-state CI in the proposed DCL framework, rather than partial CI restricted to the position-attitude subspace.

\section{WCI-Based Correlated Update}
In this section, we present the proposed WCI-based method for DCL in 3D scenarios. We begin by revisiting CI and its weighted extension, WCI. We then develop a WCI-based correlated-update approach for DCL that accounts for both the characteristics of CI fusion and the INS error propagation rule, thereby promoting balanced estimation across state components with different scales and observability levels.

\subsection{Revisiting Covariance Intersection}
Consider a state vector $\boldsymbol{x}$ with two unbiased and consistent estimates $\{\hat{\boldsymbol{x}}_1, \boldsymbol{P}_1\}$ and $\{\hat{\boldsymbol{x}}_2, \boldsymbol{P}_2\}$. Each estimate can be written as $\hat{\boldsymbol{x}}_i = \boldsymbol{x}+\delta{\boldsymbol{x}}_i$, $i=1,2$, where $\mathbb{E}[\delta{\boldsymbol{x}}_i] = \boldsymbol{0}$. We define the actual covariance matrices and the cross-covariance matrix as
\begin{equation}
    \boldsymbol{P}_i^{*} = \mathbb{E}\left[ \delta{\boldsymbol{x}}_i \delta{\boldsymbol{x}}_i^{\mathsf{T}} \right], \quad \boldsymbol{P}_{12}^{*} = \mathbb{E}\left[ \delta{\boldsymbol{x}}_1 \delta{\boldsymbol{x}}_2^{\mathsf{T}} \right].
\label{eq14}
\end{equation}
Because the matrices $\boldsymbol{P}_i^{*}$ are generally unknown, we refer to $\boldsymbol{P}_1$ and $\boldsymbol{P}_2$ as nominal covariance matrices. Consistency of each estimate then requires $\boldsymbol{P}_{i} \succeq \boldsymbol{P}_{i}^{*}$ \cite{jazwinski2013stochastic}.

CI produces an unbiased and consistent fused estimate $\{\hat{\boldsymbol{x}}, \boldsymbol{P}\}$ without requiring the cross-correlation information \cite{julier1997non}:
\begin{equation}
\begin{aligned}
    \boldsymbol{P} &= \left( \omega \boldsymbol{P}_1^{-1} + (1-\omega)  \boldsymbol{P}_2^{-1} \right)^{-1} \\
    \hat{\boldsymbol{x}} &= \boldsymbol{P} \left(
    \omega \boldsymbol{P}_1^{-1}\hat{\boldsymbol{x}}_1 + (1-\omega) \boldsymbol{P}_2^{-1} \hat{\boldsymbol{x}}_2
    \right)
\end{aligned},
\label{eq15}
\end{equation}
where $\omega \in [0,1]$ is the weight parameter. It has been shown that for any admissible $\omega$ and $\boldsymbol{P}_{12}^{*}$, the fused covariance matrix satisfies $\boldsymbol{P} \succeq \boldsymbol{P}^{*}$. The weight parameter $\omega$ is determined by minimizing a cost function $f(\boldsymbol{P};\omega)$:
\begin{equation}
\omega^{*} = \arg \min_{\omega \in [0,1]} f(\boldsymbol{P};\omega).
\label{eq16}
\end{equation}
Typically, the cost function is chosen as the trace or determinant of the fused covariance matrix, i.e., $f(\boldsymbol{P}) = \operatorname{tr}(\boldsymbol{P})$ or $\det(\boldsymbol{P})$; we refer to them as the traditional criteria.

Furthermore, for partial-observation scenarios such as $\boldsymbol{x}_2 = \boldsymbol{H}\boldsymbol{x}$, where $\boldsymbol{H}$ is rank-deficient and $\boldsymbol{P}_1^{-1}+\boldsymbol{H}^{\mathsf{T}}  \boldsymbol{P}_2^{-1} \boldsymbol{H}$ is invertible, the CI estimate is given by \cite{arambel2001covariance}:
\begin{equation}
\begin{aligned}
    \boldsymbol{P} = \left( \omega \boldsymbol{P}_1^{-1} + (1-\omega) \boldsymbol{H}^{\mathsf{T}}  \boldsymbol{P}_2^{-1} \boldsymbol{H} \right)^{-1} \\
    \hat{\boldsymbol{x}} = \boldsymbol{P} \left(
    \omega \boldsymbol{P}_1^{-1} \hat{\boldsymbol{x}}_1 + (1-\omega) \boldsymbol{H}^{\mathsf{T}} \boldsymbol{P}_2^{-1} \hat{\boldsymbol{x}}_2
    \right)
\end{aligned}.
\label{eq17}
\end{equation}

The traditional trace-based criterion follows from minimizing the MSE of the estimate, or equivalently, the trace of the actual covariance matrix (cf. \eqref{eq18}). In practice, since the actual covariance matrix is unavailable, the trace of the nominal covariance matrix is minimized instead:
\begin{equation}
    \min \mathbb{E}[(\boldsymbol{x} - \hat{\boldsymbol{x}})^{\mathsf{T}} (\boldsymbol{x} - \hat{\boldsymbol{x}})] = \operatorname{tr}(\boldsymbol{P}^{*}) \rightarrow \min \operatorname{tr}(\boldsymbol{P}).
    \label{eq18}
\end{equation}
However, minimizing MSE alone is too coarse and lacks task specificity. To address this, Arambel \textit{et al.} introduced relative weights to regulate the CI fusion process \cite{arambel2001covariance}, which we refer to as WCI. In their work, however, these weights were treated mainly as tuning knobs for improving selected state components, without a clear theoretical interpretation. We show in \eqref{eq19} that minimizing the weighted trace of the fused covariance matrix is mathematically equivalent to minimizing a WMSE over the entire state. By choosing the weighting matrix $\boldsymbol{W}$ to reflect task-dependent priorities or target WMSE levels, WCI introduces task relevance and flexibility into the fusion criterion:
\begin{equation}
    \min \mathbb{E}[(\boldsymbol{x} - \hat{\boldsymbol{x}})^{\mathsf{T}} \boldsymbol{W} (\boldsymbol{x} - \hat{\boldsymbol{x}})] = \operatorname{tr}(\boldsymbol{W}\boldsymbol{P}^{*}) \rightarrow \min \operatorname{tr}(\boldsymbol{W} \boldsymbol{P}).
    \label{eq19}
\end{equation}

Traditional CI criteria can become inadequate when the state vector contains components with substantial disparities in both scale and observability. Let $\boldsymbol{x} = [x_1^{\transpose},\ldots,x_M^{\transpose}]^{\transpose}$ denote a partitioned state vector, where different components may have different physical units, numerical magnitudes, and measurement sensitivities. The scale disparity can be reflected by large differences in the nominal block variances $\operatorname{tr}(\boldsymbol{P}_{x_m})$, whereas the observability disparity can be reflected by whether each block is directly, weakly, or only indirectly coupled to the available measurements.

Under the trace-based criterion, $f(\boldsymbol{P}) = \operatorname{tr}(\boldsymbol{P}) = \sum_i P_{ii}$, all variances are accumulated in their absolute numerical units. As a result, components with larger scales tend to dominate the optimization, whereas components with smaller scales are overshadowed and receive insufficient attention. The fusion process is therefore biased toward improving larger-scale components at the expense of inflating the uncertainty of smaller-scale components.

Under the determinant-based criterion, $f(\boldsymbol{P}) = \det(\boldsymbol{P}) = \prod_i \lambda_i(\boldsymbol{P})$, or equivalently $\log\det(\boldsymbol{P}) = \sum_i \log \lambda_i(\boldsymbol{P})$, where $\lambda_i(\boldsymbol{P})$ denotes the $i$-th eigenvalue of $\boldsymbol{P}$, the optimization depends on relative changes in the eigenvalues rather than their absolute numerical magnitudes. Therefore, unlike the trace criterion, it is not directly dominated by large-scale state components. However, it suffers from a dimensionality imbalance when the state contains an observable subspace $\mathcal{O}$ and a weakly observable or unobservable subspace $\mathcal{U}$. In CI fusion, informative measurements typically reduce the uncertainty in $\mathcal{O}$, while the consistency requirement may simultaneously inflate the uncertainty in $\mathcal{U}$. If $n_{\mathcal{U}} > n_{\mathcal{O}}$, where $n_{\mathcal{O}}$ and $n_{\mathcal{U}}$ denote the dimensions of these two subspaces, the inflation in the larger weakly observable subspace can dominate the overall volume change. More specifically, if $\boldsymbol{P}^{-}$ and $\boldsymbol{P}^{+}$ denote the covariance matrices before and after CI fusion, then
\begin{equation}
    \log \frac{\det(\boldsymbol{P}^{+})}{\det(\boldsymbol{P}^{-})}
    = \sum_{i \in \mathcal{O}} \log \frac{\lambda_i^{+}}{\lambda_i^{-}}
    + \sum_{j \in \mathcal{U}} \log \frac{\lambda_j^{+}}{\lambda_j^{-}},
    \label{eq20}
\end{equation}
where typically $\lambda_i^{+} < \lambda_i^{-}$ for observable directions and $\lambda_j^{+} > \lambda_j^{-}$ for weakly observable directions. When the second term dominates, the overall uncertainty volume increases even though the informative subspace has been improved. Since the determinant-based criterion explicitly penalizes such volume increase, it tends to suppress the contribution of the fusion observation and thus underutilize the available measurements.

\begin{figure*}[t]
    \centering
    \includegraphics[width=\linewidth]{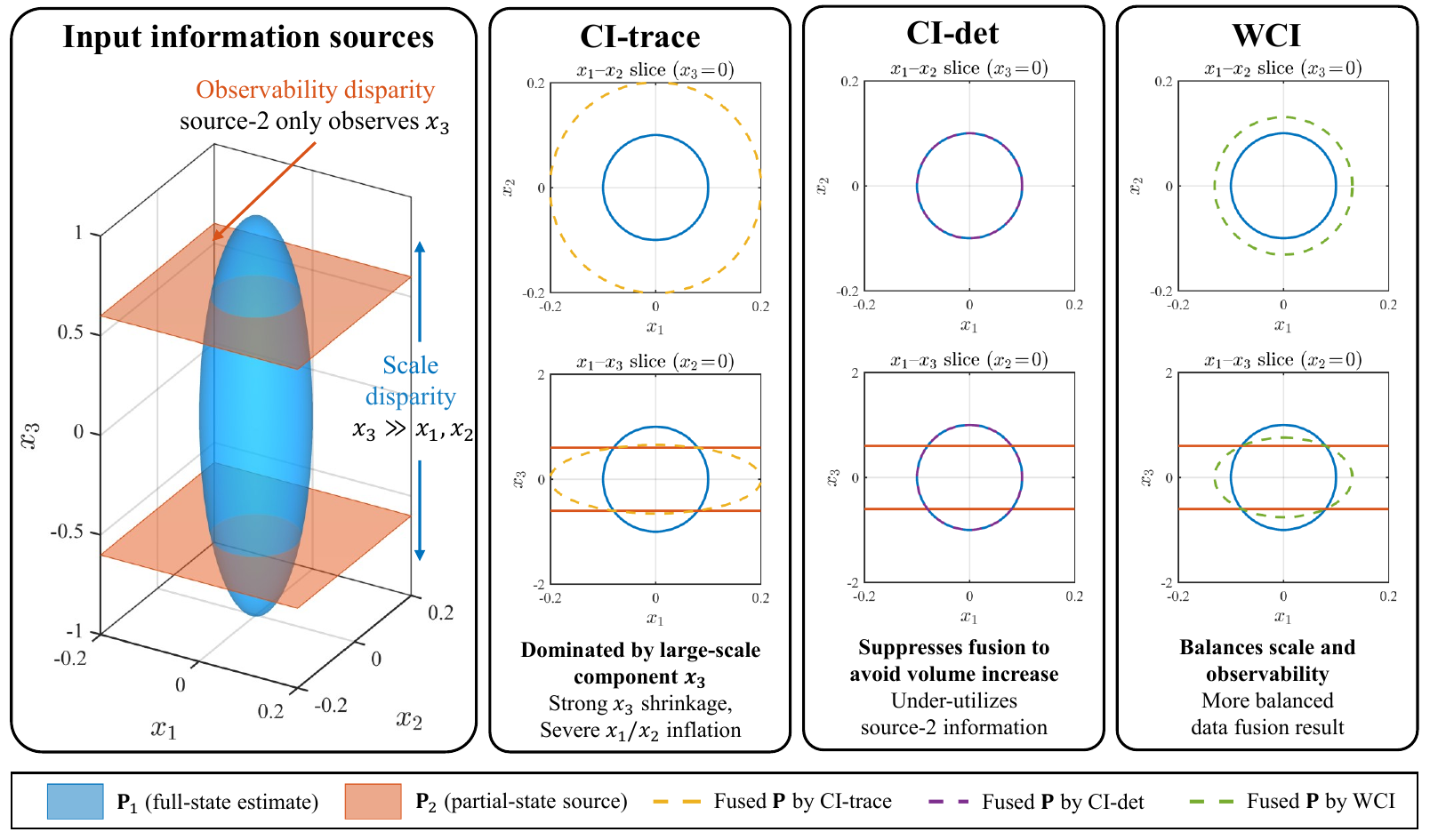}
    \caption{Illustrative three-dimensional CI example with scale and observability disparities. The prior covariance is $\boldsymbol{P}_1=\operatorname{diag}(0.1^2,0.1^2,1^2)$, and the second source observes only $x_3$ with standard deviation $0.6$. The leftmost panel shows the prior ellipsoid together with the partial observation planes, and the three panels on the right compare the fused covariances in the $x_1$--$x_2$ and $x_1$--$x_3$ slices. The $x_2$--$x_3$ slice is omitted because it is identical to the $x_1$--$x_3$ slice under the symmetric prior variances of $x_1$ and $x_2$. CI-trace overemphasizes the directly observed large-scale direction and inflates the unobservable small-scale subspace, CI-det largely suppresses the observation, and WCI achieves a more balanced fusion outcome.
    }
    \label{fig5}
\end{figure*}

To illustrate the above effect, consider a three-dimensional state $\boldsymbol{x} = [x_1,x_2,x_3]^{\transpose}$ with prior covariance
\begin{equation*}
    \boldsymbol{P}_1 = \operatorname{diag}(0.1^2,\,0.1^2,\,1^2)
\end{equation*}
and a partial-state source that only observes $x_3$, i.e.,
\begin{equation*}
    \boldsymbol{H} = \begin{bmatrix} 0 & 0 & 1 \end{bmatrix}, \qquad \boldsymbol{P}_2 = 0.6^2.
\end{equation*}
In this setting, $x_1$ and $x_2$ form an unobservable small-scale subspace with respect to the second source, whereas $x_3$ is the only directly observed large-scale component. Fig.~\ref{fig5} and Table~\ref{tab2} show the corresponding geometry and fused covariances obtained by CI-trace, CI-det, and WCI, where the percentages in Table~\ref{tab2} denote relative changes with respect to the prior covariance $\boldsymbol{P}_1$. Specifically, the leftmost panel of Fig.~\ref{fig5} visualizes the prior geometry, while the three panels on the right show how different CI criteria reshape the $x_1$--$x_2$ and $x_1$--$x_3$ uncertainty slices. CI-trace is dominated by the large-scale component $x_3$ and therefore selects a small $\omega$ to exploit the improvement in $x_3$, which reduces the variance of $x_3$ by 57.2\% but inflates the variances of $x_1$ and $x_2$ by 300\%. By contrast, CI-det nearly rejects the informative observation and yields a fused covariance close to $\boldsymbol{P}_1$, because the unobservable subspace has a larger dimension than the observable subspace and the resulting volume increase due to uncertainty inflation in the unobservable subspace dominates the volume decrease from the improvement in the observable subspace. WCI achieves a more balanced result, with a moderate 42.5\% reduction in the variance of $x_3$ and a more reasonable 70.0\% increase in the variances of $x_1$ and $x_2$, thereby leveraging information from both sources while avoiding excessive bias toward any particular component.

This issue is particularly pronounced in range-based 3D cooperative localization, where position, velocity, attitude, and IMU bias states differ substantially in both scale and observability. Therefore, the fusion criterion should explicitly account for these disparities instead of treating all state components uniformly.
\begin{table}
\centering
\caption{Fused covariance matrices for the illustrative example under different CI criteria.
}
\label{tab2}
\setlength{\tabcolsep}{4pt}
\renewcommand{\arraystretch}{1.1}
\begin{tabular}{>{\centering\arraybackslash}p{1.2cm}
                >{\centering\arraybackslash}p{0.7cm}
                >{\centering\arraybackslash}p{1.8cm}
                >{\centering\arraybackslash}p{1.8cm}
                >{\centering\arraybackslash}p{1.8cm}}
\hline
Criterion & $\omega^*$ & $P_{x_1}$ & $P_{x_2}$ & $P_{x_3}$ \\
\hline
CI-trace & 0.248 & 0.040 (+300\%) & 0.040 (+300\%) & 0.428 (-57.2\%) \\
CI-det   & 1.000 & 0.010 (0\%)    & 0.010 (0\%)    & 1.000 (0\%) \\
WCI      & 0.584 & 0.017 (+70.0\%) & 0.017 (+70.0\%) & 0.575 (-42.5\%) \\
\hline
\end{tabular}
\end{table}

\subsection{Concurrent Update Strategy}
In this subsection, building on the DCL framework introduced in Section~\textrm{III}, we present the WCI-based correlated-update process. Using vehicle $i$ as an example, assume that it has obtained differential distances to $N$ other vehicles and the corresponding covariance matrices, denoted by $\left\{ \delta d_{ij}^{\mathcal{V}}, \boldsymbol{R}_{ij}^{\mathcal{V}}  \right\}$, $j=1,\ldots,N$, whose explicit expressions are given in \eqref{eq9} and \eqref{eq11}, respectively. Since the differential measurement noise contains the estimation errors of other vehicles' states, which are correlated with the state of vehicle $i$ and whose correlations are unknown, CI, or its weighted extension WCI introduced in Section~\textrm{IV}-A, can be used to perform the correlated update.

Unlike EKF, for which sequential and concurrent updates of multiple observations yield the same result, CI does not provide the same guarantee. In fact, CI ensures that the fusion of two data sources provides an optimal conservative upper bound, but it cannot guarantee the same for the fusion of multiple data sources \cite{ajgl2018fusion}. Sequential updates of multiple observations therefore introduce additional performance degradation and yield a more conservative upper bound. Accordingly, for multiple inter-vehicle distance measurements, we combine them into a single observation vector for state update:
\begin{equation}
    \boldsymbol{z}_{i}^{\mathcal{V}} = \left[ \delta {d_{i1}^{\mathcal{V}}}, \ldots, \delta {d_{iN}^{\mathcal{V}}}   \right]^{\transpose}
    = \boldsymbol{H}_{i}^{\mathcal{V}} \delta \boldsymbol{x}_{i} + \boldsymbol{G}_{i}^{\mathcal{V}} \boldsymbol{v}_{i}^{\mathcal{V}},
\label{eq21}
\end{equation}
where the explicit expressions of $\boldsymbol{H}_{i}^{\mathcal{V}}$ and $\boldsymbol{G}_{i}^{\mathcal{V}}$ are given by
\begin{equation}
\begin{aligned}
\boldsymbol{H}_{i}^{\mathcal{V}} &= [{\boldsymbol{H}_{i1}^{\mathcal{V}}}^{\transpose}, \ldots, {\boldsymbol{H}_{iN}^{\mathcal{V}}}^{\transpose}]^{\transpose} \\
\boldsymbol{G}_{i}^{\mathcal{V}} &= [{\boldsymbol{G}_{i1}^{\mathcal{V}}}^{\transpose}, \ldots, {\boldsymbol{G}_{iN}^{\mathcal{V}}}^{\transpose}]^{\transpose} 
\end{aligned}.
\label{eq22}
\end{equation}
The combined noise vector and its actual covariance matrix can be written as
\begin{equation}
\begin{aligned}
    \boldsymbol{v}_{i}^{\mathcal{V}} &= \left[ {\boldsymbol{v}_{i1}^{\mathcal{V}}}^{\transpose}, \ldots,{\boldsymbol{v}_{iN}^{\mathcal{V}}}^{\transpose}
    \right]^{\transpose} \\
    \boldsymbol{R}_{i}^{\mathcal{V}} &= 
    \begin{bmatrix}
    \boldsymbol{R}_{i1}^{\mathcal{V}} &\cdots &\boldsymbol{C}_{i,1N}^{\mathcal{V}} \\
    \vdots &\ddots &\vdots \\
    {\boldsymbol{C}_{i,1N}^{\mathcal{V}}}^{\transpose} &\cdots &\boldsymbol{R}_{iN}^{\mathcal{V}}
\end{bmatrix}
\end{aligned}.
\label{eq23}
\end{equation}
Here, $\boldsymbol{C}_{i,1N}^{\mathcal{V}}$ denotes the cross-covariance matrix between $\boldsymbol{v}_{i1}^{\mathcal{V}}$ and $\boldsymbol{v}_{iN}^{\mathcal{V}}$, which is generally nonzero because the states of vehicle 1 and vehicle $N$ may be correlated. However, $\boldsymbol{C}_{i,1N}^{\mathcal{V}}$ is unavailable in practical distributed systems because inter-vehicle correlations are not tracked. An alternative is to appropriately inflate the noise covariance matrix while ensuring that the nominal covariance matrix remains greater than the actual covariance matrix, thereby maintaining consistency. We therefore focus on the diagonal blocks of $\boldsymbol{R}_{i}^{\mathcal{V}}$ and multiply the vehicle-state error variance terms by 2 to obtain the nominal noise covariance matrix $\boldsymbol{R}_{i}^{\mathcal{V}'}$ as follows:
\begingroup
\setlength{\arraycolsep}{4.5pt}
\begin{equation}
    \boldsymbol{R}_{i}^{\mathcal{V}'} = 
    \begin{bmatrix}
        2\boldsymbol{P}_{p_1}^{d} &\boldsymbol{0}_{3\times3} &\boldsymbol{0}_{3\times1} &\cdots &\boldsymbol{0}_{3\times3} &\boldsymbol{0}_{3\times3} &\boldsymbol{0}_{3\times1} \\
        \boldsymbol{0}_{3\times3} &2\boldsymbol{P}_{\phi_1}^{d} &\boldsymbol{0}_{3\times1} &\cdots &\boldsymbol{0}_{3\times3} &\boldsymbol{0}_{3\times3} &\boldsymbol{0}_{3\times1} \\
        \boldsymbol{0}_{1\times3} &\boldsymbol{0}_{1\times3} &\sigma_r^2 &\cdots &\boldsymbol{0}_{1\times3} &\boldsymbol{0}_{1\times3} &0 \\
        \vdots &\vdots &\vdots &\ddots &\vdots &\vdots &\vdots \\
        \boldsymbol{0}_{3\times3} &\boldsymbol{0}_{3\times3} &\boldsymbol{0}_{3\times1} &\cdots &2\boldsymbol{P}_{p_N}^{d} &\boldsymbol{0}_{3\times3} &\boldsymbol{0}_{3\times1} \\
        \boldsymbol{0}_{3\times3} &\boldsymbol{0}_{3\times3} &\boldsymbol{0}_{3\times1} &\cdots &\boldsymbol{0}_{3\times3} &2\boldsymbol{P}_{\phi_N}^{d} &\boldsymbol{0}_{3\times1} \\
        \boldsymbol{0}_{1\times3} &\boldsymbol{0}_{1\times3} &0 &\cdots &\boldsymbol{0}_{1\times3} &\boldsymbol{0}_{1\times3} &\sigma_r^2
    \end{bmatrix},
\label{eq24}
\end{equation}
\endgroup
where 
\begin{equation}
\begin{aligned}
    \boldsymbol{P}_{p_j}^{d} = \operatorname{diag}(\left[ \sigma_{p_{j,x}}^2, \sigma_{p_{j,y}}^2, \sigma_{p_{j,z}}^2  \right]^{\transpose}) \\
    \boldsymbol{P}_{\phi_j}^{d} = \operatorname{diag}(\left[ \sigma_{\phi_{j,x}}^2, \sigma_{\phi_{j,y}}^2, \sigma_{\phi_{j,z}}^2  \right]^{\transpose})
\end{aligned}
\label{eq25}
\end{equation}
($j=1,\ldots,N$) are diagonal matrices formed from the error variances of the position and attitude of vehicle $j$. It is straightforward to verify that $\boldsymbol{R}_{i}^{\mathcal{V}'} \succeq \boldsymbol{R}_{i}^{\mathcal{V}}$, which means that the nominal covariance matrix $\boldsymbol{R}_{i}^{\mathcal{V}'}$ still provides a consistent estimate of the differential measurement vector. In addition, this nominal covariance matrix is free of correlation terms, thereby eliminating the need to track correlations and reducing the communication burden.

With the inter-vehicle differential measurements $\left\{ \boldsymbol{z}_{i}^{\mathcal{V}}, \boldsymbol{R}_{i}^{\mathcal{V}'} \right\}$, CI can be applied for state update. A KF-style CI formulation for partial observations is provided in \cite{li2013split} as follows:
\begin{gather}
    \boldsymbol{P}_1 = \frac{1}{\omega}\boldsymbol{P}_{i,k|k-1} \notag \\
    \boldsymbol{P}_2 = \frac{1}{1-\omega}\boldsymbol{R}_{i}^{\mathcal{V}'} \notag \\
    \boldsymbol{K} = \boldsymbol{P}_1 {\boldsymbol{H}_{i}^{\mathcal{V}}}^{\transpose} \left( {\boldsymbol{H}_{i}^{\mathcal{V}}} \boldsymbol{P}_1 {\boldsymbol{H}_{i}^{\mathcal{V}}}^{\transpose} + \boldsymbol{G}_{i}^{\mathcal{V}} \boldsymbol{P}_2 {\boldsymbol{G}_{i}^{\mathcal{V}}}^{\transpose}
    \right)^{-1} \label{eq26} \\
    \delta{\boldsymbol{x}}_{i,k} = \boldsymbol{K} \boldsymbol{z}_{i}^{\mathcal{V}} \notag \\
    \boldsymbol{P}_{i,k} = (\boldsymbol{I} - \boldsymbol{K}{\boldsymbol{H}_{i}^{\mathcal{V}}}) \boldsymbol{P}_1 \notag.
\end{gather}
Here, $\omega$ is the weight parameter, which is determined by a specific optimization criterion in \eqref{eq16}.

In range-based 3D cooperative localization, position is typically the directly observed large-scale component, whereas attitude and IMU biases are relatively small-scale and weakly observable, with velocity lying between these two extremes through INS coupling. Accordingly, CI-trace tends to overemphasize position and inflate the remaining states, whereas CI-det may suppress informative observations when weakly observable directions dominate. Therefore, we adopt WCI in \eqref{eq19} to achieve more balanced full-state fusion, and the remaining design issue is the selection of the weighting matrix.

\subsection{Weighting Matrix Selection}
\begin{figure*}[!t]
\hrulefill
\begin{equation}
    \boldsymbol{S} = \begin{bmatrix}
    \boldsymbol{I}_3,\  \boldsymbol{I}_3 \cdot T_a,\  
    \frac{1}{2} \left[ \left(\boldsymbol{C}_{b_i}^{n}\boldsymbol{f}_{i}^{b} \right) \times \right]  T_a^2,\
    -\frac{1}{6} \left[ \left(\boldsymbol{C}_{b_i}^{n}\boldsymbol{f}_{i}^{b} \right) \times \right]\boldsymbol{C}_{b_i}^{n} T_a^3,\ 
    \frac{1}{2}\boldsymbol{C}_{b_i}^{n} T_a^2
    \end{bmatrix}.
\tag{30}
\label{eq30}
\end{equation}

\begin{equation}
    \mathbb{E} \left[  
    \left(\prescript{I}{}{\delta{\boldsymbol{p}}}_{i,T_a}^{n}\right)^{\transpose}  \left(\prescript{I}{}{\delta{\boldsymbol{p}}}_{i,T_a}^{n}\right)
    \right]
    = \operatorname{tr}\left( \mathbb{E} \left[ 
    \left(\prescript{I}{}{\delta{\boldsymbol{p}}}_{i,T_a}^{n}\right)  \left(\prescript{I}{}{\delta{\boldsymbol{p}}}_{i,T_a}^{n}\right)^{\transpose}
    \right] \right)
    = \operatorname{tr}\left( \boldsymbol{S}
    \mathbb{E} \left[ 
    \delta\boldsymbol{x}_{i,0} \delta\boldsymbol{x}_{i,0}^{\transpose}
    \right] \boldsymbol{S}^{\transpose}
    \right)
    = \operatorname{tr}\left( \boldsymbol{S}^{\transpose}\boldsymbol{S}
    \boldsymbol{P}_{i,0}
    \right).
\tag{31}
\label{eq31}
\end{equation}
\hrulefill
\end{figure*}

Inertial navigation serves as the primary mechanism for state propagation in vehicle-state estimation, whereas distance measurements provide supplementary correction. Reducing inertial-navigation error can therefore lower the overall cooperative-localization error and improve autonomous navigation when distance measurements are insufficient. Since inertial-navigation error depends on all state components, this objective also provides guidance for balancing their relative importance in the CI weighting matrix.

We now consider the propagation law of inertial-navigation errors. Since vehicles are generally equipped with low-cost MEMS sensors, we neglect factors such as the Earth's rotation, the variation of frame $\mathcal{N}$, and changes in gravity. Under these assumptions, a simplified error-state model can be expressed as \cite{niu2021wheel}
\begin{equation}
\left\{
\begin{aligned}
\prescript{I}{}{\delta{\dot{\boldsymbol{p}}}_{i}^{n}} &= \prescript{I}{}{\delta{\boldsymbol{v}}_{i}^{n}} \\
\prescript{I}{}{\delta{\dot{\boldsymbol{v}}}_{i}^{n}} &= \boldsymbol{f}_{i}^{n}\times \boldsymbol{\phi}_i^n + \boldsymbol{C}_{b_i}^{n} \delta\boldsymbol{f}_{i}^{b} \\
\dot{\boldsymbol{\phi}}_i^n &= - \boldsymbol{C}_{b_i}^{n} \delta\boldsymbol{\omega}_{i}^{b}
\end{aligned}
\right.,
\label{eq27}
\end{equation}
where $\boldsymbol{f}_{i}^{n} = \boldsymbol{C}_{b_i}^{n} \boldsymbol{f}_{i}^{b}$ is the representation of the specific force $\boldsymbol{f}_{i}^b$ in frame $\mathcal{N}$. The terms $\delta \boldsymbol{f}_{i}^{b}$ and $\delta\boldsymbol{\omega}_{i}^{b}$ represent the errors in the specific-force and angular-velocity outputs of the IMU, respectively. These errors are modeled as
\setcounter{equation}{27}
\begin{equation}
\begin{aligned}
    \delta \boldsymbol{f}_{i}^{b} = \boldsymbol{b \scriptstyle a}_i + \boldsymbol{w}_{a} \\
    \delta\boldsymbol{\omega}_{i}^{b} = \boldsymbol{b \scriptstyle g}_i + \boldsymbol{w}_{g}
\end{aligned},
\label{eq28}
\end{equation}
where $\boldsymbol{w}_{a}$ and $\boldsymbol{w}_{g}$ are zero-mean random noise terms. This error-state model shows that inertial-navigation position error accumulates from different state components with different time orders: the initial position error contributes as a zero-order term, the initial velocity error as a first-order term, the initial attitude error and accelerometer bias as second-order terms, and the initial gyroscope bias as a third-order term \cite{titterton2004strapdown}. This hierarchical propagation structure suggests that different state components should not be treated uniformly when designing the CI weighting matrix.

To make this idea explicit, consider a simplified motion segment in which vehicle $i$ moves at constant velocity along a straight line over a time interval of length $T_a$. The inertial-navigation position error at time $T_a$ can then be approximated as
\begin{equation}
\begin{split}
    \prescript{I}{}{\delta{\boldsymbol{p}}}_{i,T_a}^{n} =\ & \prescript{I}{}{\delta{\boldsymbol{p}}}_{i,0}^{n} + \prescript{I}{}{\delta{\boldsymbol{v}}}_{i,0}^{n} T_a + 
    \frac{1}{2} \left[ \left(\boldsymbol{C}_{b_i}^{n} \boldsymbol{f}_{i}^{b} \right) \times \right] \boldsymbol{\phi}_{i}^{n}  T_a^2 + \\ 
    &\frac{1}{2}\boldsymbol{C}_{b_i}^{n}\boldsymbol{b \scriptstyle a}_i T_a^2 - 
    \frac{1}{6} \left[ \left(\boldsymbol{C}_{b_i}^{n} \boldsymbol{f}_{i}^{b} \right) \times \right]\boldsymbol{C}_{b_i}^{n} \boldsymbol{b \scriptstyle g}_i T_a^3 \\
   =\ & \boldsymbol{S} \delta \boldsymbol{x}_{i,0},
\end{split}
\label{eq29}
\end{equation}
where the explicit expression for $\boldsymbol{S}$ is given by \eqref{eq30}. We then use the mean squared inertial-navigation position error at time $T_a$ as the design objective. According to \eqref{eq31}, minimizing this quantity is equivalent to minimizing the trace of $\boldsymbol{S}^{\transpose}\boldsymbol{S}\boldsymbol{P}_{i,0}$. Therefore, choosing the weighting matrix as in \eqref{eq32} makes the optimization objective in \eqref{eq31} exactly consistent with the WMSE criterion in \eqref{eq19}.
\setcounter{equation}{31}
\begin{equation}
    \boldsymbol{W} = \boldsymbol{S}^{\transpose} \boldsymbol{S},
\label{eq32}
\end{equation}

In practice, however, vehicles do not strictly maintain straight-line constant-velocity motion, and the exact matrix $\boldsymbol{S}$ in \eqref{eq30} depends on the instantaneous specific force and attitude at the update time. To reduce this dependence on local maneuvering conditions while preserving the dominant propagation-order information, we replace $\boldsymbol{S}$ with the following simplified approximation, which retains only the time-order scaling of each state component:
\begin{equation}
    \tilde{\boldsymbol{S}} = \begin{bmatrix}
    \boldsymbol{I}_3 &
    T_a \boldsymbol{I}_3 &
    T_a^2 \boldsymbol{I}_3 &
    T_a^3 \boldsymbol{I}_3 &
    T_a^2 \boldsymbol{I}_3
    \end{bmatrix}.
    \label{eq33}
\end{equation}
This form follows directly from \eqref{eq29}, where position, velocity, attitude, gyroscope bias, and accelerometer bias contribute to the position error with orders $0$, $1$, $2$, $3$, and $2$, respectively. Although $\tilde{\boldsymbol{S}}$ is not an exact surrogate for the motion-dependent matrix in \eqref{eq30}, it preserves the dominant propagation-order priorities while avoiding dependence on instantaneous maneuvers and attitude. We therefore use $\boldsymbol{W}_{i,k}=\tilde{\boldsymbol{S}}^{\transpose}\tilde{\boldsymbol{S}}$ for online fusion:
\begin{equation}
    \boldsymbol{W}_{i,k} = \tilde{\boldsymbol{S}}^{\transpose}\tilde{\boldsymbol{S}}.
    \label{eq34}
\end{equation}

With $\boldsymbol{W}_{i,k}$ defined above, the weight parameter $\omega$ in \eqref{eq26} is determined by
\begin{equation}
\omega^{*} = \arg \min_{\omega \in [0,1]} \operatorname{tr}(\boldsymbol{W}_{i,k}\boldsymbol{P}_{i,k}),
\label{eq35}
\end{equation}
where the estimated error state $\delta \boldsymbol{x}_{i,k}$ is then used to correct the nominal state estimate. The complete WCI-based correlated-update procedure is summarized in \hyperref[algorithm]{\textbf{Algorithm}}.
\begin{algorithm}[!b]
\renewcommand{\thealgorithm}{}
\caption{WCI-based correlated update procedure.}
\begin{algorithmic}[]
\Require 
\Statex
\begin{itemize}
    \item State estimate after prediction $\{ \hat{\boldsymbol{x}}_{i,k|k-1} , \boldsymbol{P}_{i,k|k-1} \}$
    \item Distance measurements to other vehicles $\{ \tilde{d}_{ij}^{\mathcal{V}}
    \}$
    \item Other vehicles' position and attitude estimates $\prescript{I}{}{\hat{\boldsymbol{p}}_{j,k|k-1}^{n}}$ and $\hat{\boldsymbol{q}}_{j,k|k-1}^{n}$, along with their nominal variances
    $\{ \sigma_{p_{j,x}}^2,\sigma_{p_{j,y}}^2, \sigma_{p_{j,z}}^2\}$ and $\{ \sigma_{\phi_{j,x}}^2,\sigma_{\phi_{j,y}}^2, \sigma_{\phi_{j,z}}^2\}$.
\end{itemize}

\Ensure State estimate after update $\{ \hat{\boldsymbol{x}}_{i,k} , \boldsymbol{P}_{i,k} \}$
\State Calculate differential measurements with the inflated nominal covariance matrices $\{ \boldsymbol{z}_{i}^{\mathcal{V}}, \boldsymbol{R}_{i}^{\mathcal{V}'} \}$ according to \eqref{eq9}--\eqref{eq12} and \eqref{eq21}--\eqref{eq25}.
\State Determine the simplified matrix $\tilde{\boldsymbol{S}}$ by \eqref{eq33} and construct $\boldsymbol{W}_{i,k}$ by \eqref{eq34}.
\State Execute data fusion using WCI by \eqref{eq26} and \eqref{eq35}.
\State Perform error feedback to correct the predicted state estimate.
\end{algorithmic}
\label{algorithm}
\end{algorithm}

The optimality of \eqref{eq35} should be interpreted with respect to the selected weighted covariance objective. In the general WCI formulation, for a fixed positive-semidefinite weighting matrix, the weighted trace is a monotone covariance cost; when the weighting matrix is positive definite, this monotonicity becomes strict. Therefore, the CI optimal-bound result under unknown correlations can be applied to the corresponding WMSE-related criterion \cite{reinhardt2015minimum}. Thus, WCI preserves the consistency-oriented interpretation of CI while introducing an additional degree of freedom to make the fusion criterion task-dependent. This flexibility is particularly useful in 3D DCL, where different state components have different physical scales, observability levels, and contributions to subsequent navigation performance. In this work, the weighting matrix is constructed from the dominant INS error-propagation orders, so that the CI fusion weight is selected according to how different state components affect subsequent inertial-navigation position error. The proposed weighting strategy is not claimed to be a universal weighting matrix that guarantees the optimal performance for every state component and every possible scenario; rather, it provides a physically interpretable and practically implementable criterion for balanced full-state estimation in the considered DCL task.

From an implementation perspective, this weighting design does not introduce additional numerical difficulty. The weighting matrix in \eqref{eq34} is positive semidefinite by construction, and its entries are controlled only by the design parameter $T_a$. The online weight selection in \eqref{eq35} therefore remains a scalar optimization over $\omega \in [0,1]$, without dependence on instantaneous maneuver-dependent coefficients. The same principle can also be extended to higher-dimensional state vectors, provided that the weighting matrix is designed according to the propagation characteristics, scales, and observability of the additional states.

In summary, the independent and correlated update procedures for vehicle-to-anchor and inter-vehicle distance measurements are handled by EKF and WCI, respectively. When both types of measurements are available, they can be processed sequentially. The updated state and covariance matrix are then passed to the next prediction step.

\section{Simulation Results}
In this section, we present simulation results to validate the proposed WCI method against traditional CI criteria in 3D cooperative localization. We also compare the proposed DCL framework with the CCL framework to highlight the robustness, computational efficiency, and communication efficiency offered by the distributed architecture.

\begin{figure}[!b]
    \centering
    \includegraphics[width=\linewidth]{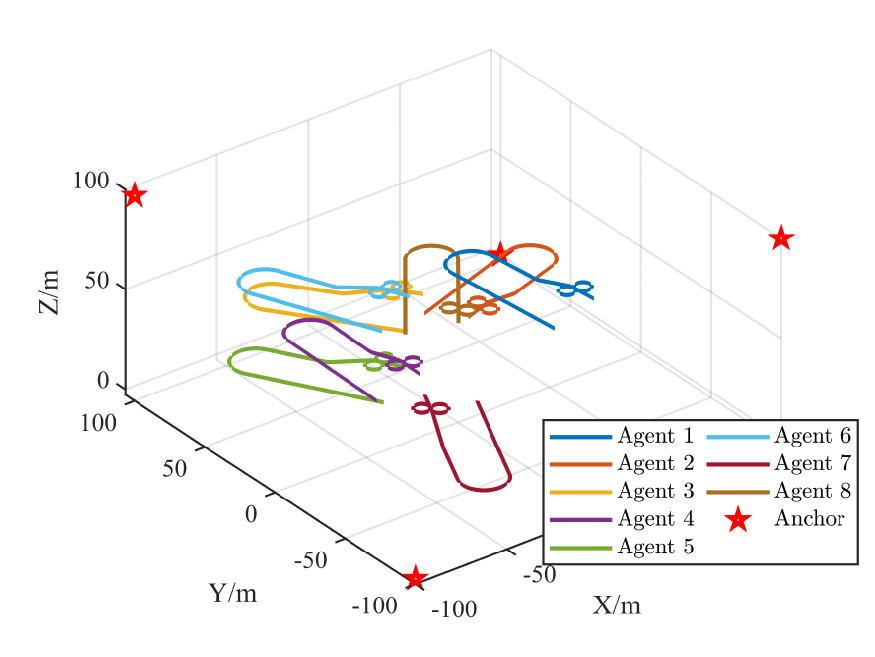}
    \caption{Anchor layout and vehicle trajectories in the simulation scenario.}
    \label{fig6}
\end{figure}

Consider a 3D cooperative localization system in which all vehicles are initially static, with random positions and random yaw angles, while roll and pitch are initialized to zero. The vehicles then execute the following motion sequence:
\begin{itemize}
\item Accelerate at \(0.4 \, \text{m/s}^2\) for 10~s.
\item Follow a figure-eight trajectory, composed of two circular motions in opposite directions, for 26~s.
\item Increase the pitch angle by $15^{\circ}$ over 1~s.
\item Move at a constant velocity for 10~s.
\item Decrease the pitch angle by $15^{\circ}$ over 1~s.
\item Move at a constant velocity for 13~s.
\item Perform a $180^{\circ}$ clockwise circular turn over 20~s.
\item Move at a constant velocity for 29~s.
\item Decelerate linearly to a hover over 10~s.
\end{itemize}
The total motion lasts 120~s, covers a variety of representative maneuvers, and sufficiently excites the observability of all state components. To prevent the vehicle states from diverging, we deploy four anchors with known positions at $[100, 100, 0]^{\transpose}~\text{m}$, $[-100, 100, 100]^{\transpose}~\text{m}$, $[-100, -100, 0]^{\transpose}~\text{m}$, and $[100, -100, 100]^{\transpose}~\text{m}$. The IMU parameters follow those of the ADIS16465 \cite{analog2020adis16465}, a representative high-precision MEMS IMU. For ranging, we adopt UWB, which can achieve centimeter-level accuracy under line-of-sight conditions \cite{malajner2015uwb}. Unless otherwise specified, the remaining simulation parameters are listed in Table~\ref{tab3}. The vehicle trajectories and anchor layout are shown in Fig.~\ref{fig6}.

The anchor connectivity of the vehicles is configured to span a spectrum of practical scenarios with varying degrees of reliance on inter-vehicle cooperation. Specifically, one vehicle is assigned connections to all four anchors, representing the best anchor-connected case where anchor-based updates are abundant; one vehicle has no anchor connections at all, representing the most challenging localization scenario in which state correction relies entirely on inter-vehicle cooperation; and the remaining vehicles each connect to two of the four anchors, reflecting a common practical scenario in which communication range limitations or physical obstructions prevent full anchor coverage. Two anchors are insufficient for accurate localization in 3D space, and thus these vehicles must rely on both anchor-based updates and inter-vehicle measurements for state estimation. This diversity in anchor connectivity allows us to evaluate how different algorithms perform under varying levels of anchor availability and cooperative dependence.

\begin{table}[!t]
    \centering
    \caption{Default simulation parameters.}
    \setlength{\tabcolsep}{3pt}
    \begin{tabular}{>{\centering\arraybackslash}p{0.13\columnwidth}
    >{\centering\arraybackslash}p{0.20\columnwidth}
    >{\raggedright\arraybackslash}p{0.55\columnwidth}}
        \toprule
        \textbf{Notation} & \textbf{Value} & \multicolumn{1}{c}{\textbf{Description}} \\
        \midrule
        $N_v$ & 8 & Number of vehicles \\
        $f_{\text{range}}$ &\(5 \, \text{Hz}\)& Range measurement frequency \\
        $\sigma_r$ & \(0.1 \, \text{m}\) & Standard deviation of range measurements \\
        $\boldsymbol{l}^{b}$ & $[0.1,0,0.1]^{\transpose}\,\text{m}$ & Lever arm from IMU to ranging module in frame $\mathcal{B}$ \\ 
        $\sigma_{p_0}$ & \(0.3 \, \text{m}\) & Standard deviation of each vehicle's initial position \\
        $\sigma_{v_0}$ & \(0.1 \, \text{m/s}\) & Standard deviation of each vehicle's initial velocity \\
        $\sigma_{\phi_0}$ &  $3^{\circ}$ & Standard deviation of each vehicle's initial attitude \\
        $T_a$ &  \(10 \, \text{s}\) & Hyperparameter of WCI \\
        $N_{\text{sim}}$ &  100 & Total number of simulation runs \\
        \bottomrule
    \end{tabular}
    \label{tab3}
\end{table}

We compare the proposed WCI method with the following four benchmarks.
\begin{enumerate}
    \item \textbf{NCL}: No cooperative localization, where each vehicle corrects its state using only distance measurements from anchors. This method serves as a baseline to highlight the impact of inter-vehicle cooperation.
    \item \textbf{CCL EKF}: A CCL method in which the computing center\footnote{Unless otherwise specified, we assume that all IMU and distance measurements can be transmitted to the computing center without packet loss.} maintains a large covariance matrix for all vehicles' states and utilizes EKF for updates using all available measurements. Since it tracks the correlations between vehicles, it is considered the performance ceiling for cooperative localization.
    \item \textbf{DCL CI-trace}: A DCL method in which each vehicle runs its own filter and uses trace-based CI for updates with measurements between vehicles.
    \item \textbf{DCL CI-det}: A DCL method in which each vehicle runs its own filter and uses determinant-based CI for updates with measurements between vehicles.
\end{enumerate}

The evaluation metrics include root-mean-squared error (RMSE), root-mean-trace error (RMTE) \cite{chang2021resilient}, and normalized estimation error squared (NEES) \cite{barshalom2001estimation}. 
RMSE reflects the actual estimation error, whereas RMTE characterizes the uncertainty represented by the nominal covariance matrix. Consistency requires the nominal covariance to be no smaller than the actual estimation-error covariance. Beyond this basic requirement, we further expect the nominal covariance to predict the actual estimation error as closely as possible. A smaller RMTE is desirable only when consistency is still satisfied, because it then indicates a tighter uncertainty bound. Intuitively, the ratio $\text{RMSE}/\text{RMTE}$ provides a useful indicator of estimator tightness: values much smaller than one indicate an overly conservative estimator, whereas values greater than one indicate an overconfident and potentially inconsistent estimator. In addition, NEES is a widely used metric for evaluating estimator consistency because it measures the estimation error relative to its covariance matrix. Ideally, for a random vector of dimension $n$, the theoretical NEES value is $n$; if NEES is significantly larger or smaller than $n$, the estimator is excessively optimistic or pessimistic, respectively.
Let $\boldsymbol{y}$ denote the state component under consideration. At time step $k$ in Monte Carlo run $n$, the state estimate and true value are denoted by $\hat{\boldsymbol{y}}_k^{(n)}$ and $\boldsymbol{y}_k^{(n)}$, respectively, with the nominal covariance matrix given by $\boldsymbol{P}_{y,k}^{(n)}$. The formulas for these three metrics are as follows:
\begin{equation}
    \operatorname{RMSE}_{y,k} = \sqrt{\frac{1}{N_{\text{sim}}}\sum_{n=1}^{N_{\text{sim}}} (\hat{\boldsymbol{y}}_k^{(n)} - \boldsymbol{y}_k^{(n)})^{\transpose} (\hat{\boldsymbol{y}}_k^{(n)} - \boldsymbol{y}_k^{(n)})},
\label{eq36}
\end{equation}
\begin{equation}
    \operatorname{RMTE}_{y,k} = \sqrt{ \frac{1}{N_{\text{sim}}}\sum_{n=1}^{N_{\text{sim}}} \operatorname{tr}(\boldsymbol{P}_{y,k}^{(n)})},
\label{eq37}
\end{equation}
\begin{equation}
    \operatorname{NEES}_{y,k} = \frac{1}{N_{\text{sim}}}\sum_{n=1}^{N_{\text{sim}}} (\hat{\boldsymbol{y}}_k^{(n)} - \boldsymbol{y}_k^{(n)})^{\transpose} (\boldsymbol{P}_{y,k}^{(n)})^{-1} (\hat{\boldsymbol{y}}_k^{(n)} - \boldsymbol{y}_k^{(n)}),
\label{eq38}
\end{equation}
where $N_{\text{sim}}$ is the total number of simulation runs.

\subsection{DCL Analysis}
\begin{figure}[!b]
\centering
\subfloat[]{\includegraphics[width=\linewidth]{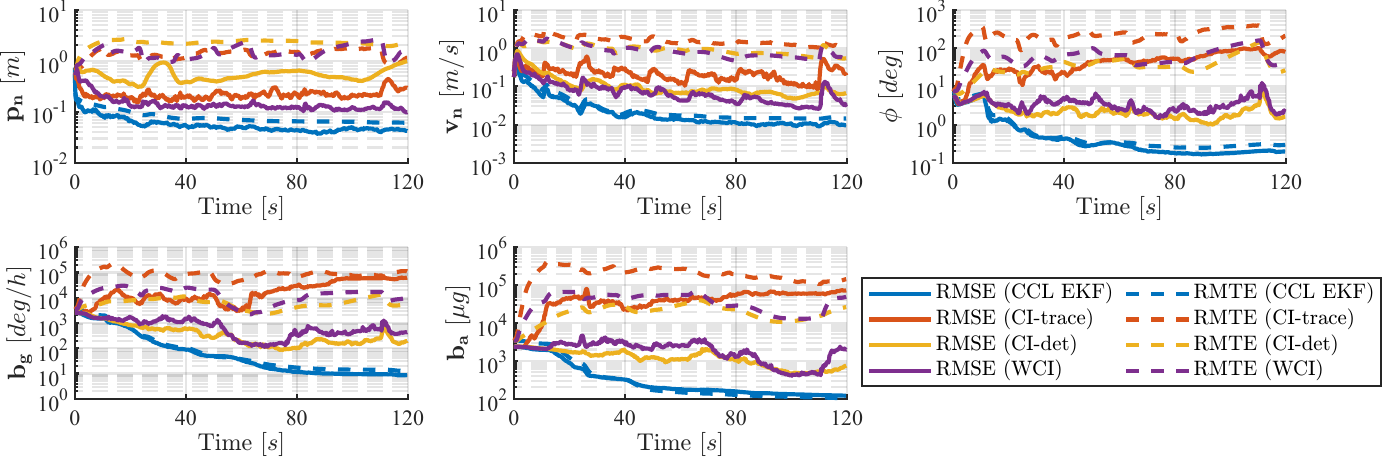}%
\label{fig7_1}}
\hfil
\subfloat[]{\includegraphics[width=\linewidth]{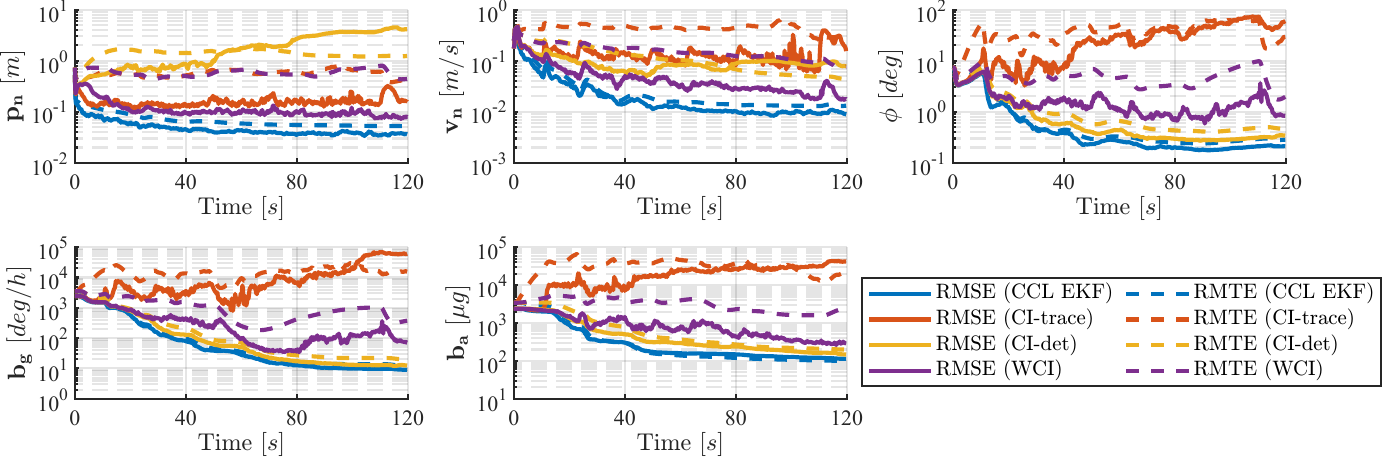}%
\label{fig7_2}}
\hfil
\subfloat[]{\includegraphics[width=\linewidth]{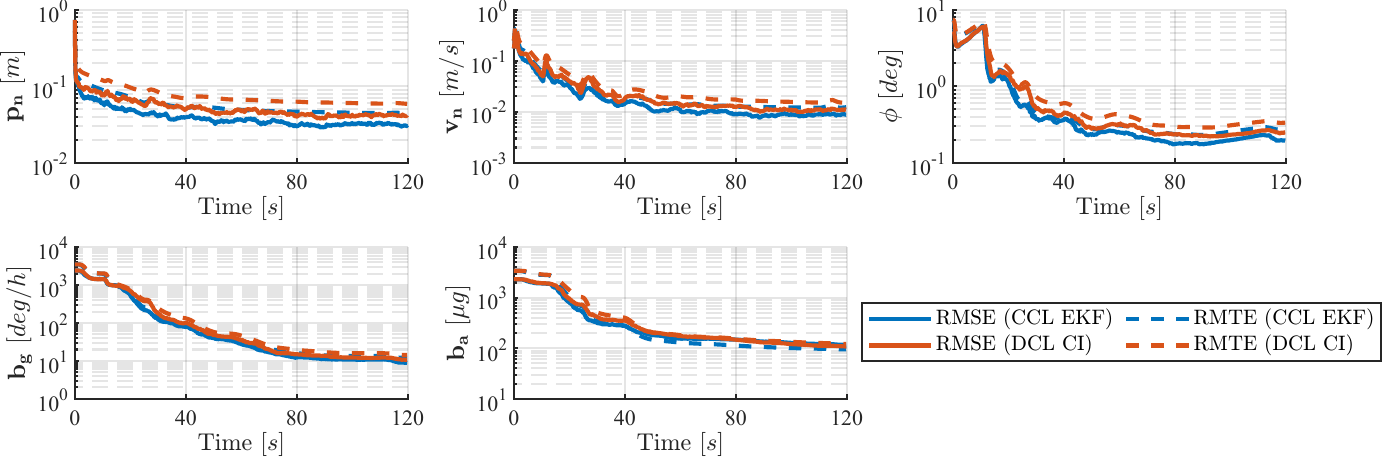}%
\label{fig7_3}}
\caption{RMSE and RMTE of all state components for different methods. (a) Vehicle 1. (b) Vehicle 7. (c) Vehicle 8.}
\label{fig7}
\end{figure}

In this subsection, we analyze the performance of the proposed WCI-based DCL algorithm against all benchmarks, with particular emphasis on the comparison with traditional CI criteria. The simulation parameters are set to their default values. 
To represent the three anchor-connectivity scenarios, we select one vehicle from each category: Vehicle 1 for the no-anchor-connectivity case, Vehicle 7 as a representative of the partial-anchor-connectivity case, and Vehicle 8 for the full-anchor-connectivity case.
The RMSE and RMTE for all state components under different methods are shown in Fig.~\ref{fig7}. The RMSE, RMTE, and NEES values averaged over all time steps for the three representative vehicles are reported in Table~\ref{tab4}\footnote{Since Vehicle 8 has full anchor connectivity, the state update is dominated by anchor-based EKF updates, and the contribution of inter-vehicle CI fusion is negligible. Hence, CI-trace, CI-det, and WCI yield nearly identical results, and only one representative result is reported as DCL CI.}.

\begin{table*}[!ht]
    \centering
    \caption{RMSE, RMTE, and NEES of all state components for different methods, averaged over all time steps.}
    \setlength{\tabcolsep}{1.5pt}
    \captionsetup[subfloat]{font=normal}
    \subfloat[RMSE, RMTE, and NEES for Vehicle 1 with no anchor connectivity.]{%
        \begin{tabular}{l *{15}{>{\raggedright\arraybackslash}p{0.95cm}}}
        \toprule
        \multirow{2}{*}{\textbf{Method}} 
        & \multicolumn{3}{c}{\textbf{Position [m]}} 
        & \multicolumn{3}{c}{\textbf{Velocity [m/s]}} 
        & \multicolumn{3}{c}{\textbf{Attitude [deg]}} 
        & \multicolumn{3}{c}{\textbf{Gyro Bias [deg/h]}} 
        & \multicolumn{3}{c}{\textbf{Accel Bias [$\boldsymbol{\mu}$g]}} \\
        \cmidrule(lr){2-4} \cmidrule(lr){5-7} \cmidrule(lr){8-10} \cmidrule(lr){11-13} \cmidrule(lr){14-16}
        & RMSE & RMTE & NEES & RMSE & RMTE & NEES & RMSE & RMTE & NEES & RMSE & RMTE & NEES & RMSE & RMTE & NEES \\
        \midrule
        \multicolumn{1}{l}{\text{$\quad$NCL}}  & 4030& 3967 & 25.11 & 118.6 & 118.1 & 50.11 & 29.25 & 28.39 & 3.041 & 2507 & 2494 & 3.031 & 2359 & 2425 & 2.840 \\
        \multicolumn{1}{l}{\text{CCL EKF}}  & 0.059 & 0.059 & 2.911 & 0.031 & 0.028 & 3.506 & 0.790 & 0.701 & 4.123 & 291.2 & 261.0 & 4.169 & 495.1 & 450.1 & 5.70 \\
        \multicolumn{1}{l}{\text{DCL CI-trace}} & 0.235 & 1.047 & 0.271 & 0.237 & 1.176 & 0.209 & 45.87 & 157.0 & 2.804 & 2.3E+04 & 6.2E+04 & 1.055 & 4.3E+04 & 1.5E+05 & 1.020 \\
        \multicolumn{1}{l}{\text{DCL CI-det}} & 0.546 & 1.643 & 0.301 &
        0.121 & 0.672 & 0.147 & 2.367 & 27.43 & 0.146 & 522.9 & 4583 & 0.148 & 1274 & 1.4E+04 & 0.151 \\
        \multicolumn{1}{l}{\text{DCL WCI}} & 0.170 & 1.152 & 0.158  & 0.083 & 0.519 & 0.109 & 1.837 & 19.83 & 0.129 & 466.7 & 3848  & 0.128 & 1002  & 1.1E+04 & 0.122 \\
        \bottomrule
    \end{tabular}
    }    
    \vspace{0.0cm} 
    \subfloat[RMSE, RMTE, and NEES for Vehicle 7 with partial anchor connectivity.]{%
        \begin{tabular}{l *{15}{>{\raggedright\arraybackslash}p{0.95cm}}}
        \toprule
        \multirow{2}{*}{\textbf{Method}} 
        & \multicolumn{3}{c}{\textbf{Position [m]}} 
        & \multicolumn{3}{c}{\textbf{Velocity [m/s]}} 
        & \multicolumn{3}{c}{\textbf{Attitude [deg]}} 
        & \multicolumn{3}{c}{\textbf{Gyro Bias [deg/h]}} 
        & \multicolumn{3}{c}{\textbf{Accel Bias [$\boldsymbol{\mu}$g]}} \\
        \cmidrule(lr){2-4} \cmidrule(lr){5-7} \cmidrule(lr){8-10} \cmidrule(lr){11-13} \cmidrule(lr){14-16}
        & RMSE & RMTE & NEES & RMSE & RMTE & NEES & RMSE & RMTE & NEES & RMSE & RMTE & NEES & RMSE & RMTE & NEES \\
        \midrule
        \multicolumn{1}{l}{\text{$\quad$NCL}}  & 9.023 & 2.085 
        & 1.4E+04 & 0.300 & 0.141 & 92.44 & 1.403 & 0.921 & 14.07     
        & 353.9 & 302.9 & 10.18 & 898.7 &  715.6 & 23.58 \\
        \multicolumn{1}{l}{\text{CCL EKF}}  & 0.051 & 0.051 & 2.913  & 0.028 & 0.026 & 3.527 &  0.777 & 0.683 & 4.129 & 284.2
        & 257.1 & 4.314 & 490.3 & 437.4 & 5.607 \\
        \multicolumn{1}{l}{\text{DCL CI-trace}} & 0.168 & 0.423      
        & 2.479 & 0.151 & 0.322 & 4.417 & 28.27 & 27.02 & 16.39    
        & 1.4E+04 & 1.0E+04 & 7.924 & 2.0E+04 & 2.3E+04 & 7.805 \\
        \multicolumn{1}{l}{\text{DCL CI-det}} & 1.957 & 0.994 &  67.778
        & 0.097 & 0.091 & 3.350 & 0.991 & 0.960 & 3.058 & 330.5      
        & 315.9 & 3.118 & 646.9 & 682.1 & 3.286 \\
        \multicolumn{1}{l}{\text{DCL WCI}} & 0.133 & 0.469 & 1.639 & 0.049 & 0.091 & 1.365 & 1.122 & 1.631 & 1.312 & 346.0      
        & 437.4 & 1.312 & 670.6 & 1143.0 & 0.820 \\
        \bottomrule
    \end{tabular}
    }
    \vspace{0.0cm}
    \centering
    \subfloat[RMSE, RMTE, and NEES for Vehicle 8 with full anchor connectivity.]{%
        \begin{tabular}{>{\raggedright\arraybackslash}p{1.6cm} *{15}{>{\raggedright\arraybackslash}p{0.95cm}}}
        \toprule
        \multirow{2}{*}{\textbf{Method}} 
        & \multicolumn{3}{c}{\textbf{Position [m]}} 
        & \multicolumn{3}{c}{\textbf{Velocity [m/s]}} 
        & \multicolumn{3}{c}{\textbf{Attitude [deg]}} 
        & \multicolumn{3}{c}{\textbf{Gyro Bias [deg/h]}} 
        & \multicolumn{3}{c}{\textbf{Accel Bias [$\boldsymbol{\mu}$g]}} \\
        \cmidrule(lr){2-4} \cmidrule(lr){5-7} \cmidrule(lr){8-10} \cmidrule(lr){11-13} \cmidrule(lr){14-16}
        & RMSE & RMTE & NEES & RMSE & RMTE & NEES & RMSE & RMTE & NEES & RMSE & RMTE & NEES & RMSE & RMTE & NEES \\
        \midrule
        \multicolumn{1}{l}{\text{$\quad$NCL}}  & 0.056 & 0.057 
        & 2.886 & 0.030 & 0.031 & 3.200 & 0.865 & 0.789 & 3.575     
        & 297.5 & 280.9 & 3.577 & 506.2 & 489.7 & 4.102 \\
        \multicolumn{1}{l}{\text{CCL EKF}}  & 0.042 & 0.043 & 2.951 & 0.025 & 0.023 & 3.594 & 0.770 & 0.665 & 4.054 & 273.7
        & 253.0 & 4.089 & 468.5 & 422.9 & 6.138 \\
        \multicolumn{1}{l}{\text{DCL CI}} & 0.056 & 0.057 
        & 2.886 & 0.030 & 0.031 & 3.200 & 0.865 & 0.789 & 3.575     
        & 297.5 & 280.9 & 3.577 & 506.2 & 489.7 & 4.102 \\
        \bottomrule
    \end{tabular}
    }
\label{tab4}
\end{table*}

Compared with the non-cooperative localization method, all cooperative localization algorithms substantially improve state estimation, confirming that inter-vehicle cooperation is essential when anchor measurements are insufficient. 
All CI-based DCL methods exhibit RMTE values that are generally larger than the corresponding RMSE values, which is consistent with the conservative nature of CI: the nominal covariance is intended to upper-bound the actual estimation error rather than match it tightly. From Vehicle 1 to Vehicle 7 and then to Vehicle 8, the number of available anchor measurements increases. As a result, EKF-based independent updates play an increasingly dominant role in state correction, while reliance on CI-based inter-vehicle fusion decreases. Accordingly, the gap between RMTE and RMSE generally narrows as anchor connectivity improves, indicating that the estimator becomes less conservative when more anchor information is available. Since the values in Table~\ref{tab4} are averaged over the full trajectory, however, they mainly reflect overall trends rather than instantaneous estimator tightness. Full-interval consistency should therefore be interpreted together with NEES and, when necessary, the time histories in Fig.~\ref{fig7}.

Among the distributed methods, CI-trace and CI-det exhibit two opposite deficiencies. 
CI-trace achieves a relatively small position RMSE and the smallest position RMTE among the DCL methods, but this advantage is obtained at the expense of the smaller-scale states, especially attitude and IMU biases, whose errors become excessively large.
In contrast, CI-det yields better attitude and IMU-bias estimates, but performs worst in position estimation. By reweighting the fusion objective according to INS error propagation, WCI reduces the overall error level while maintaining more even performance across state components.

\begin{figure}
    \centering
    \includegraphics[width=0.95\linewidth]{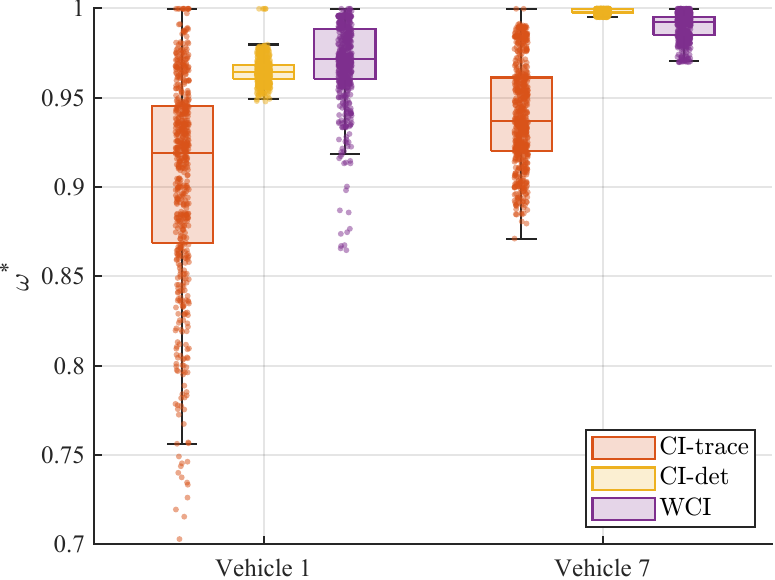}  
    \caption{CI fusion weight $\omega^*$ for Vehicles 1 and 7 under different DCL methods. CI-trace tends to choose smaller $\omega^*$ values and thus performs more aggressive range fusion, CI-det tends to choose larger $\omega^*$ values and may suppress fusion in the partial-anchor-connectivity case, while WCI stays between them and achieves a more balanced compromise.}
    \label{fig8}
\end{figure}

These state-estimation characteristics are consistent with the CI fusion-weight distributions shown in Fig.~\ref{fig8}. For Vehicles 1 and 7, CI-trace generally selects smaller $\omega^*$ values, indicating more aggressive use of inter-vehicle range information and stronger emphasis on the large-scale, directly observable position states. As a result, CI-trace favors position improvement but tends to inflate the uncertainty of the weakly observable smaller-scale states. By contrast, in the present DCL task the effectively observable subspace is dominated by position; although attitude is also coupled to the range measurements through the lever arm, its observability is much weaker. The observable subspace is therefore lower-dimensional than the weakly observable or unobservable subspace. Under the determinant-based criterion, this dimensional imbalance makes the inflation of uncertainty in the larger weakly observable subspace more influential on the overall covariance volume. CI-det therefore tends to choose larger $\omega^*$ values and suppress fusion, especially for Vehicle 7 with partial anchor connectivity, where inter-vehicle measurements play a more important role in state correction. This yields more restrained position correction while better preserving the estimates of attitude and IMU biases. WCI remains between these two extremes and adjusts the fusion strength more adaptively, thereby achieving a better overall trade-off across the full state.

\begin{figure}
    \centering
    \includegraphics[width=0.9\linewidth]{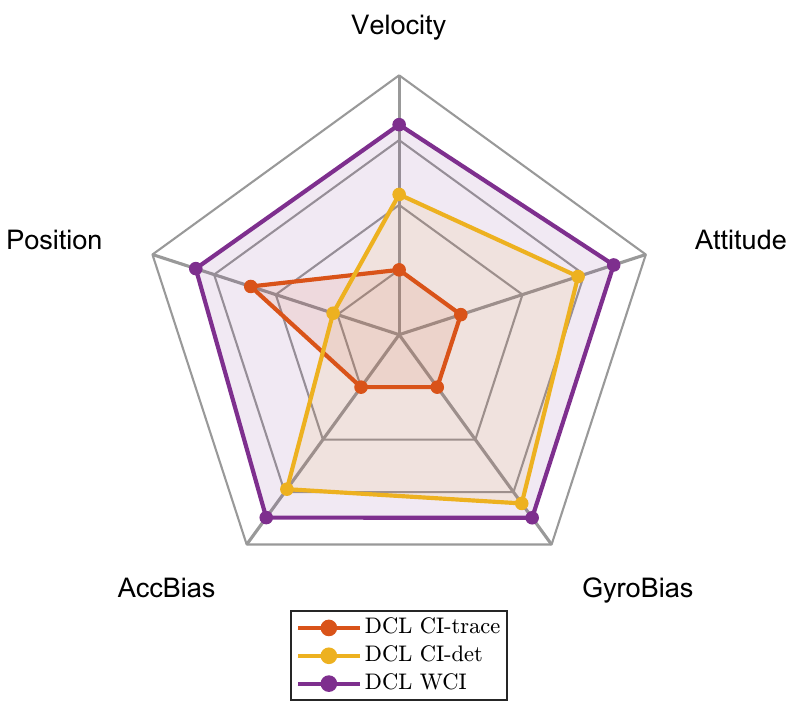}  
    \caption{Radar chart of state-estimation capability for different DCL methods on Vehicle 1. The DCL WCI capability values entirely encompass those of the traditional CI-based methods across all five state components, indicating a more balanced estimation capability.}
    \label{fig9}
\end{figure}

To compare the state-estimation capabilities of different DCL methods more directly, we use the reciprocal of the averaged RMSE of each state component as the capability metric. Fig.~\ref{fig9} presents the resulting radar chart for Vehicle 1, and similar trends are observed for the other vehicles. The figure shows that DCL WCI achieves a substantially more balanced capability profile than the traditional CI-based methods, with capability values that entirely encompass those of CI-trace and CI-det across all five state components.

The NEES results are broadly consistent with the RMSE-RMTE trend above, but they more directly reflect how well the reported nominal covariance matches the actual estimation error. Since each state component considered here is three-dimensional, the reference NEES value is 3. 
From this perspective, the DCL CI-series methods are generally conservative, as their NEES values mostly remain below 3. By contrast, the CCL EKF results are generally close to 3, but some states, especially attitude and IMU biases, still exceed 3 because first-order linearization introduces covariance mismatch in this nonlinear system. This effect becomes more pronounced for Vehicle 7 when different CI criteria are combined with EKF updates. Although the combination of CI with EKF remains consistent in principle, CI-trace and CI-det enlarge different state errors, namely attitude and IMU biases for CI-trace and position for CI-det. These enlarged errors further amplify the linearization mismatch already present in EKF, making the corresponding nominal covariance overly optimistic relative to the actual error and thus yielding larger NEES values. 
By contrast, WCI better balances fusion across heterogeneous state components and produces a more uniformly consistent and conservative estimator.

\begin{figure}
    \centering
    \includegraphics[width=0.95\linewidth]{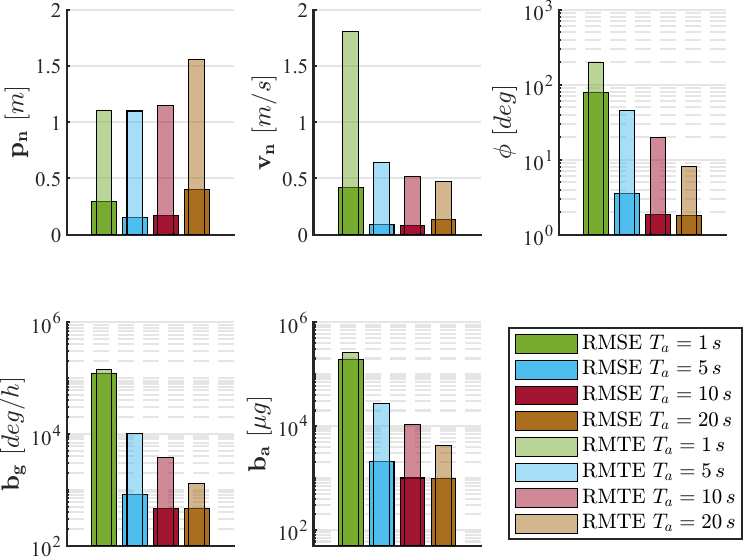} 
    \caption{State-estimation performance of DCL WCI for different values of $T_a$. As $T_a$ increases, the weighting matrix places more emphasis on states with higher-order contributions to inertial error propagation, leading to a clearer trade-off in RMTE across state components.}
    \label{fig10}
\end{figure}

We also examine the effect of the hyperparameter $T_a$ in the weighting matrix of DCL WCI. Fig.~\ref{fig10} shows the performance of DCL WCI obtained with different $T_a$ values in \eqref{eq33} and \eqref{eq34}. Since WCI optimizes the fused nominal covariance rather than the actual estimation error directly, the influence of $T_a$ is reflected more clearly in RMTE than in RMSE. A larger $T_a$ places greater emphasis on long-term inertial-navigation performance during optimization and therefore assigns more weight to the states that make higher-order contributions to propagation error, particularly attitude and IMU biases. Accordingly, as $T_a$ increases, the position RMTE gradually increases, whereas the RMTEs of attitude and IMU biases decrease more systematically.

\begin{figure}
    \centering
    \includegraphics[width=0.95\linewidth]{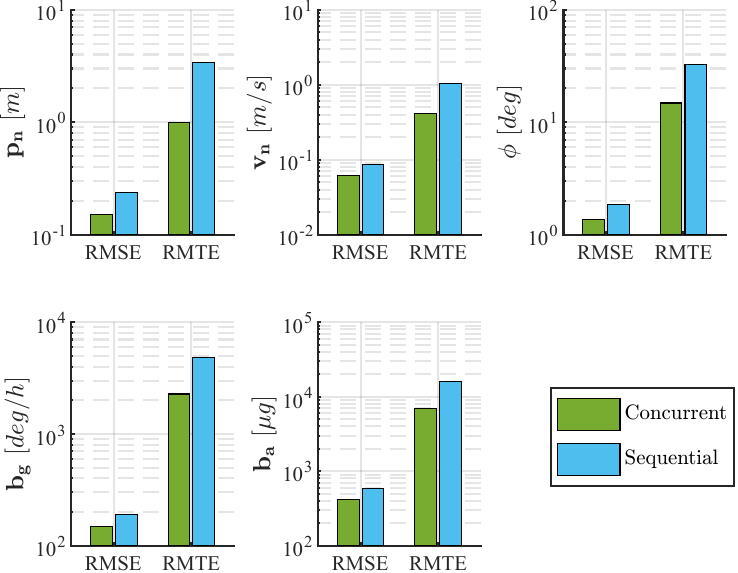}
    \caption{Comparison of concurrent and sequential update schemes. Concurrent updates yield lower conservativeness and better estimation performance than sequential updates.}
    \label{fig11}
\end{figure}

Finally, we evaluate the proposed concurrent update strategy by comparing it with its sequential counterpart, in which each vehicle updates its state immediately after receiving a single measurement. As shown in Fig.~\ref{fig11}, the concurrent strategy better matches the characteristics of CI fusion and therefore yields less conservative estimates and better estimation performance than sequential updates.

\subsection{Comparison Between DCL and CCL}
\label{sec:comparison}
In this subsection, we compare the DCL and CCL algorithms in terms of estimation accuracy, robustness, and computational and communication overhead. These three aspects reflect, respectively, the performance cost of distribution, the ability to handle packet loss and topology changes, and the scalability of each architecture in large-scale systems.

\subsubsection{Accuracy}
Compared with CCL EKF, DCL WCI does not track inter-vehicle correlations and therefore loses some correlation information during fusion, which leads to reduced estimation accuracy. According to Table~\ref{tab4}, for the two representative vehicles considered here, namely Vehicles 1 and 7, the position RMSE of DCL WCI is approximately $2.6$--$2.9\times$ that of CCL EKF, while the attitude RMSE is approximately $1.4$--$2.3\times$. Nevertheless, this accuracy remains sufficient for typical multi-vehicle-system tasks under general operating conditions, and the loss is a reasonable trade-off for the substantial gains in robustness and scalability shown below.

\subsubsection{Robustness}
\begin{figure*}[!t]
    \centering
    \includegraphics[width=\linewidth]{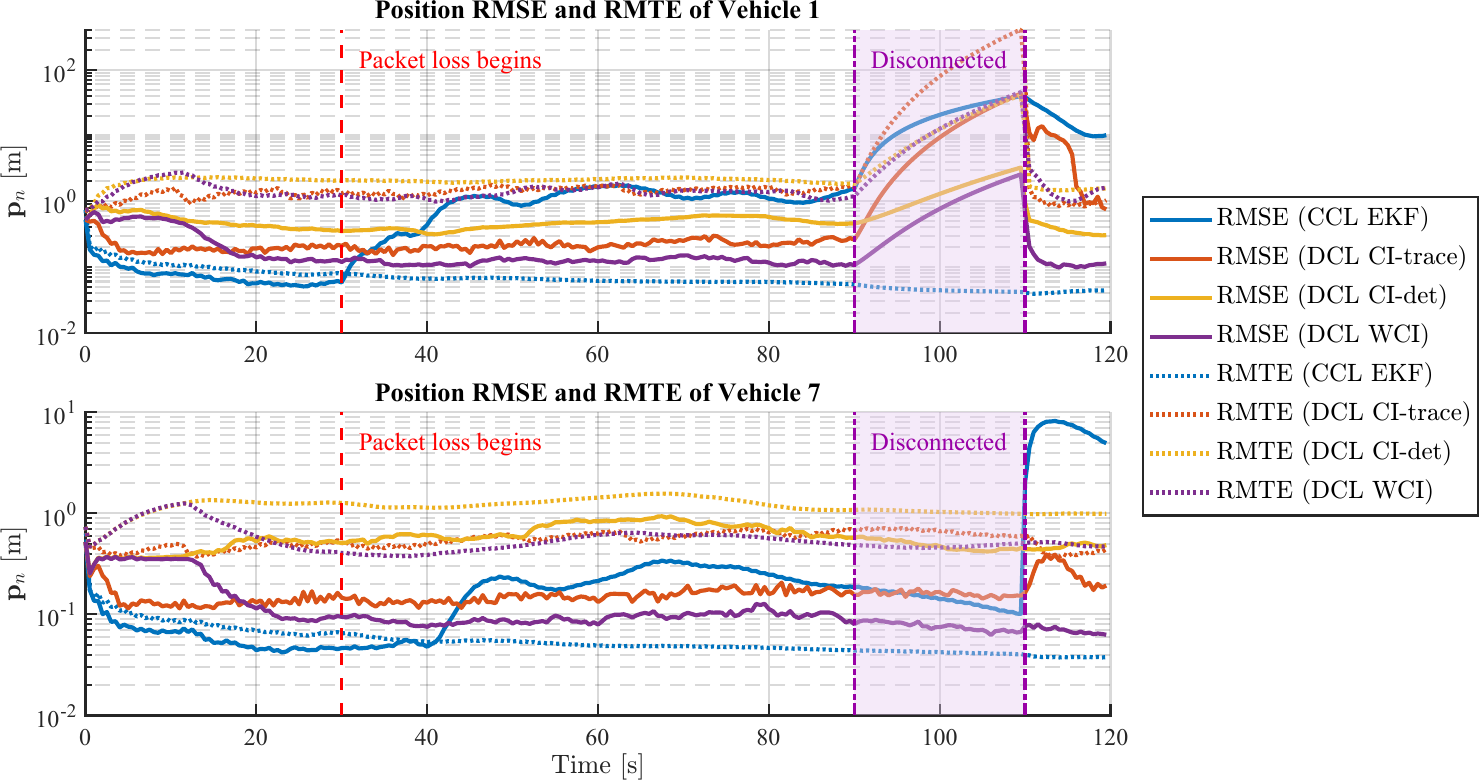} 
    \caption{Position error comparison between DCL and CCL under packet loss and topology changes. After 30 s, vehicle 1 experiences a 5\% probability of data transmission failure to the computing center. At 90 s, vehicle 1 goes offline and disconnects from both the computing center and the other vehicles. At 110 s, it rejoins the network and restores its connections to the other vehicles. DCL methods remain unaffected by these issues, while CCL EKF experiences significant degradation in position estimation, which demonstrates the superior robustness of DCL in handling communication failures and dynamic network conditions.}
    \label{fig12}
\end{figure*}

CCL requires all information to be collected at the computing center. Consequently, packet loss or the failure of a single vehicle not only interrupts the estimation of the affected vehicle but also corrupts the maintenance of inter-vehicle correlations, gradually degrading the state estimates of other vehicles. To evaluate robustness, we compare the centralized and distributed algorithms under two non-ideal conditions: 1) communication packet loss, where data from certain vehicles fails to reach the computing center, and 2) topology changes, where vehicles leave or rejoin the network. Specifically, starting from 30~s, data transmitted from Vehicle 1 to the computing center is dropped with a probability of 5\%. At 90~s, Vehicle 1 goes offline and loses its connections to both the computing center and the other vehicles. It then rejoins the network at 110~s and restores its previous connections.

We select Vehicles 1 and 7 as representative examples. As shown in Fig.~\ref{fig12}, after 30~s the position-estimation error of Vehicle 1 under CCL EKF increases gradually, and this degradation also propagates to other vehicles, such as Vehicle 7, even though they do not themselves experience packet loss. During this period, the actual RMSE exceeds the nominal RMTE, indicating a loss of consistency. By contrast, the DCL methods remain unaffected because they do not rely on globally maintained inter-vehicle correlations.

During the interval from 90~s to 110~s, Vehicle 1 is completely offline and disconnected from all anchors and other vehicles, so its state is propagated only by its own IMU outputs. Its position error therefore grows rapidly. Nevertheless, because DCL WCI provides more balanced estimates of attitude and IMU biases before disconnection, it accumulates error more slowly during this autonomous-navigation phase. After reconnection, the centralized algorithm still suffers from the stale state of Vehicle 1 accumulated during the offline interval. Consequently, the differential distance measurements associated with Vehicle 1 become significantly biased and further distort the position estimates of other vehicles, such as Vehicle 7. In contrast, the DCL framework keeps estimation local to each vehicle and is therefore substantially more robust to packet loss and topology changes.

\subsubsection{Computation and Communication Overhead}
\begin{figure}
    \centering
    \includegraphics[width=0.9\linewidth]{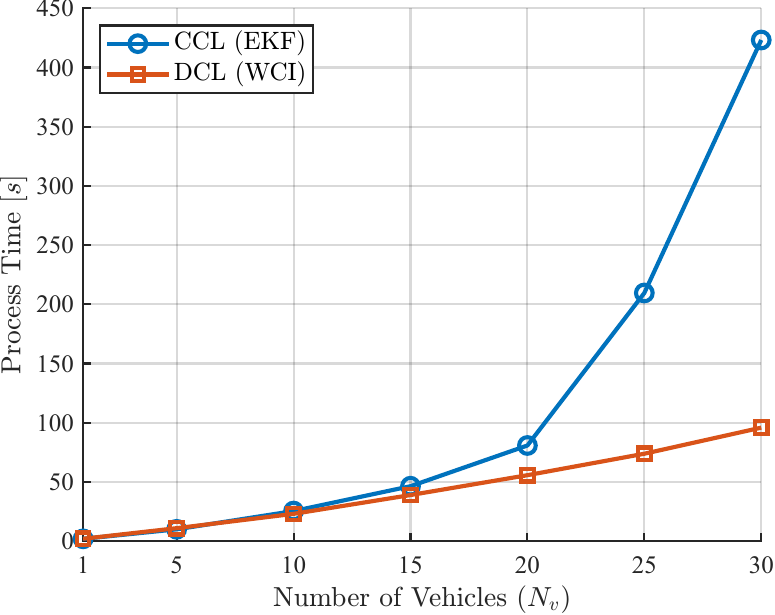} 
    \caption{The total processing time of the DCL and CCL algorithms versus swarm size. The processing time of CCL grows rapidly because the dimension of the joint covariance matrix increases with swarm size, whereas DCL exhibits approximately linear growth and is therefore more suitable for large-scale swarms.}
    \label{fig13}
\end{figure}

To evaluate computational complexity, we compare the total processing time of DCL and CCL under different swarm sizes on the same platform (MATLAB R2022b, Intel Core i7-12700, 32-GB RAM), as shown in Fig.~\ref{fig13}. For DCL, the total computational cost increases approximately linearly with the number of vehicles. Moreover, when averaged over all vehicles, the per-vehicle computational load for local state estimation remains nearly constant as the swarm size grows, which makes the distributed architecture well suited to large-scale systems. By contrast, in CCL the dimension of the joint covariance matrix grows linearly with the number of vehicles, so the cost of the matrix multiplications and inversions involved in prediction and update grows cubically with the state dimension \cite{zhang2017matrix}. Consequently, the total processing time of CCL increases rapidly with swarm size, making it difficult to scale to large vehicle networks.

The communication burden shows a similar contrast. CCL requires high-rate IMU and measurement data from all vehicles to be transmitted to the computing center, and the vehicles must in turn receive their updated state estimates from the center. In some distributed implementations of CCL, vehicles additionally broadcast state estimates and state-transition matrices to support correlation tracking \cite{li2025distributed}, which further increases the communication load. By contrast, the proposed DCL framework requires each vehicle to broadcast only its position and attitude estimates together with the associated variances, amounting to only a few dozen bytes per message. The communication requirement is therefore substantially lower. Since this conclusion follows directly from the information exchanged by the two architectures, we do not conduct separate simulation experiments for this aspect.

\section{Conclusion}
In this paper, we proposed a WCI-based DCL algorithm for MVS operating in 3D scenarios. Unlike existing studies that treat CI mainly as a plug-in fusion method, we focused on improving the CI-based data-fusion process for 3D cooperative localization. We showed that traditional CI criteria do not achieve satisfactory performance in this problem because of substantial disparities in scale and observability among different state components. To address this issue, we introduced WCI into range-based DCL in 3D scenarios and designed both a concurrent multi-measurement fusion strategy tailored to the characteristics of CI and a weighting-matrix selection strategy guided by INS error-propagation principles. Simulation results show that the proposed DCL WCI algorithm outperforms DCL methods based on traditional CI criteria in 3D cooperative localization and achieves more balanced state estimation across heterogeneous state components. The proposed concurrent fusion strategy yields higher precision and tighter uncertainty bounds than sequential updates. In addition, the distributed framework demonstrates stronger robustness and better scalability, which makes it well suited to large-scale MVS.

Future work will focus on developing dynamic weighting-matrix selection strategies that adapt to changes in vehicle motion and cluster configurations, 
validating the proposed method with real-world experimental data under non-ideal communication conditions,
and exploring more advanced distributed data-fusion methods to further reduce the conservativeness of CI.

\bibliographystyle{IEEEtran}
\bibliography{IEEEabrv,reference}

\begin{IEEEbiography}[{\includegraphics[width=1in,height=1.25in,clip,keepaspectratio]{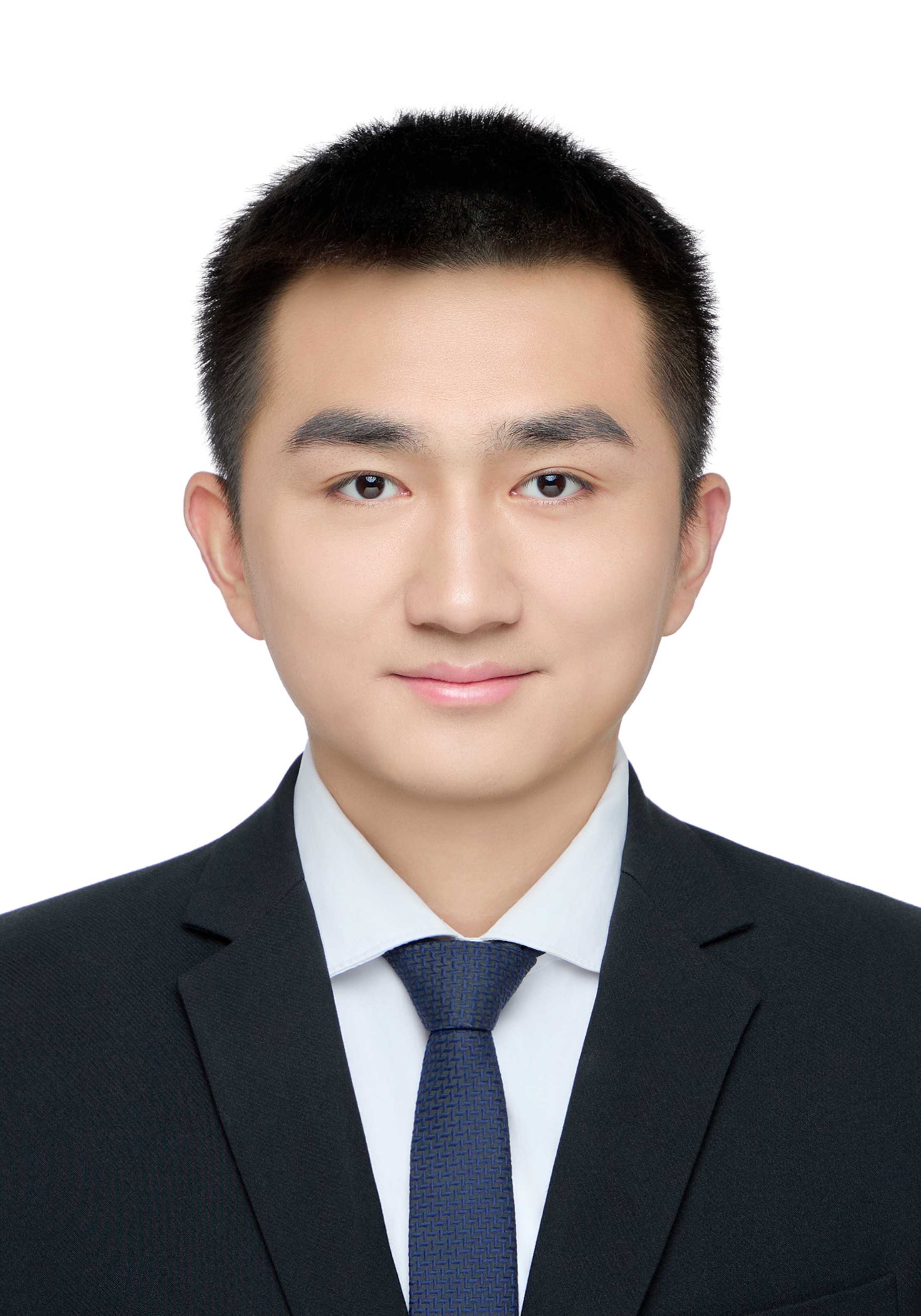}}]{Chenxin Tu} received the B.E. degree in electronic engineering from Tsinghua University, Beijing, China, in 2022, where he is currently pursuing the Ph.D. degree with the Department of Electronic Engineering.
His research interests include wireless localization and cooperative localization.
\end{IEEEbiography}

\begin{IEEEbiography}[{\includegraphics[width=1in,height=1.25in,clip,keepaspectratio]{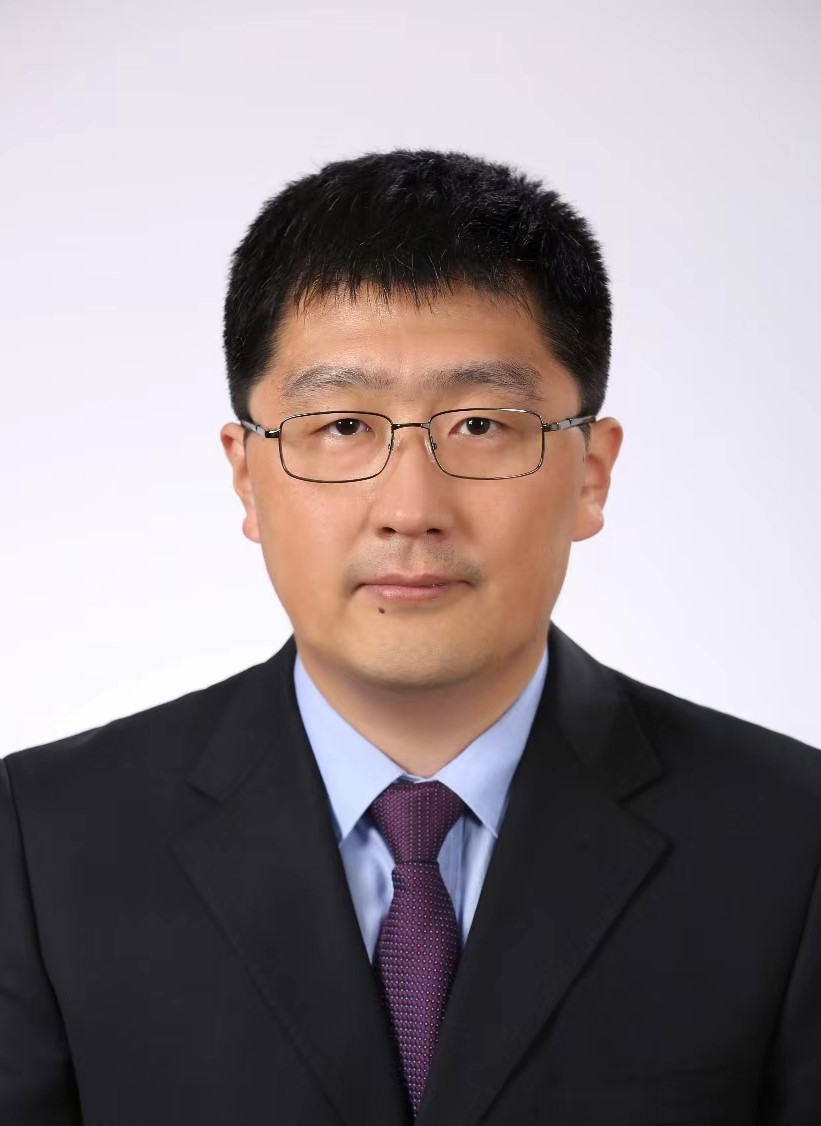}}]{Xiaowei Cui} received the B.S. and Ph.D. degrees in electronic engineering from Tsinghua University, Beijing, China, in 2000 and 2005, respectively. Since 2005, he has been with the Department of Electronic Engineering, Tsinghua University, where he is currently a professor.

His research interests include robust GNSS signal processing, multipath mitigation techniques, and high-precision positioning. He is a member of the Expert Group of China BeiDou Navigation Satellite System.
\end{IEEEbiography}

\begin{IEEEbiography}[{\includegraphics[width=1in,height=1.25in,clip,keepaspectratio]{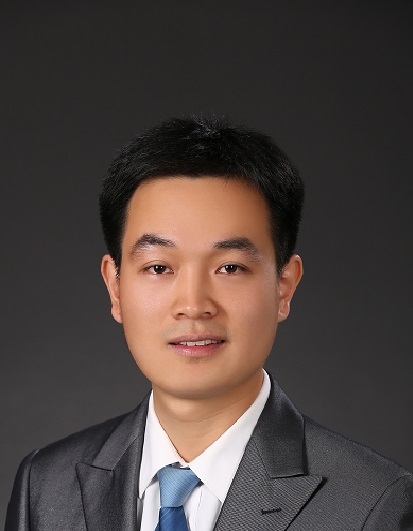}}]{Gang Liu}
received the B.S. and Ph.D. degrees in instrument science and technology from Tsinghua University, Beijing, China, in 2007 and 2015, respectively. He is currently an associate researcher with the Department of Electronic Engineering, Tsinghua University. His research interests include GNSS/INS integrated navigation and high-precision localization.
\end{IEEEbiography}

\begin{IEEEbiography}[{\includegraphics[width=1in,height=1.25in,clip,keepaspectratio]{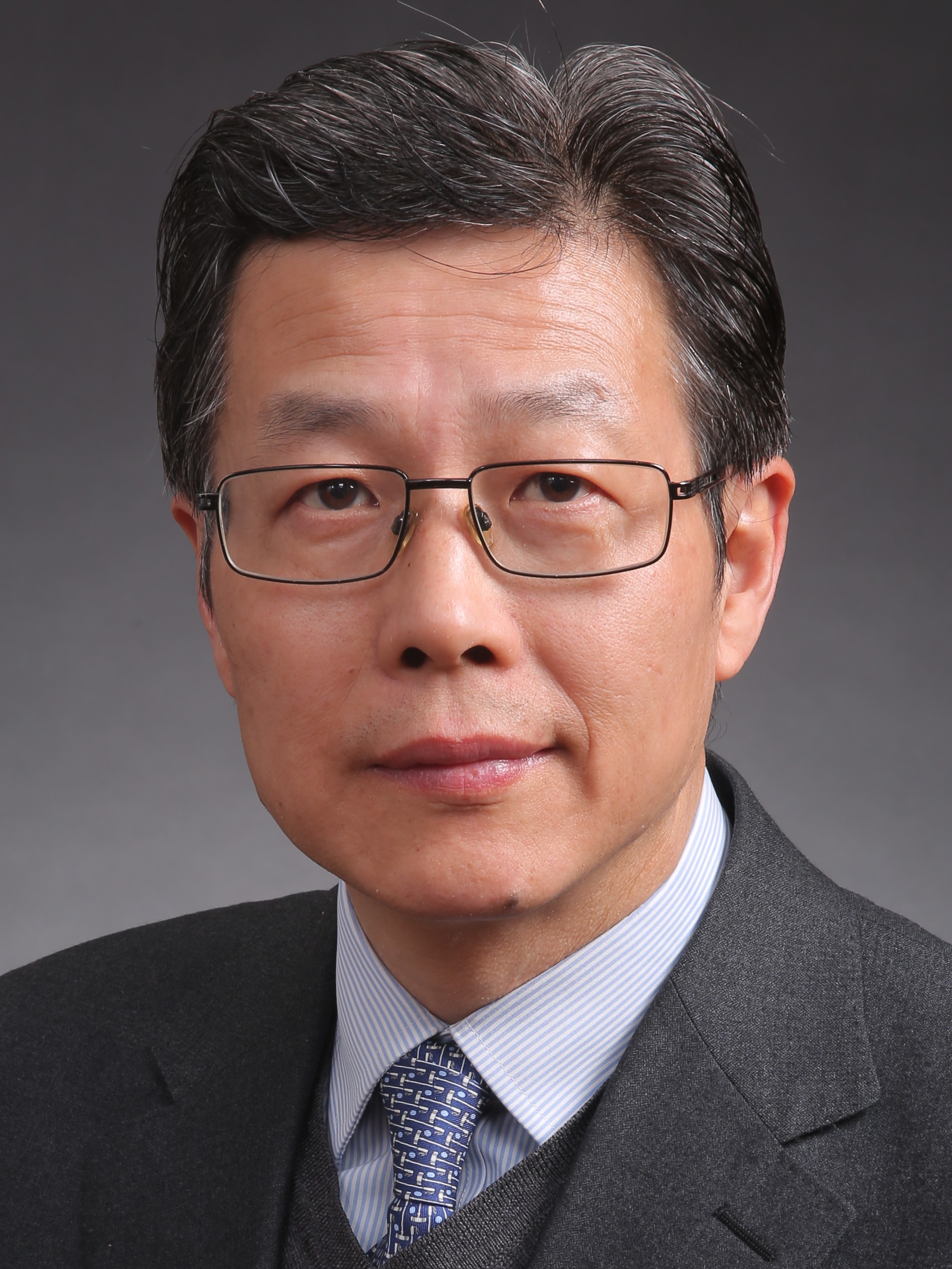}}]{Mingquan Lu}
received the M.S. and Ph.D. degrees from University of Electronic Science and Technology of China, Chengdu, China. He is currently a professor with the Department of Electronic Engineering, Tsinghua University. He directs the PNT Research Center, which develops GNSS and alternative PNT technologies. His current research focuses on GNSS signal processing, GNSS receiver development, and emerging PNT technologies. He has authored or co-authored five books and book chapters, published more than 300 journal and conference papers, and holds nearly 100 patents. He has served the GNSS community in numerous roles, including as an associate editor of \textit{Satellite Navigation} and the \textit{Journal of Navigation and Positioning}. He is a Fellow of the ION and a recipient of the ION Thurlow Award.
\end{IEEEbiography}

\end{document}